\documentclass{emulateapj}
\usepackage[usenames, dvipsnames]{color}

\usepackage{natbib} 
\usepackage{multirow}
\usepackage{threeparttable}  
\usepackage[most]{tcolorbox}
\usepackage{hyperref}
\usepackage{xcolor}
\hypersetup{
  colorlinks   = true, 
  urlcolor     = blue, 
  linkcolor    = blue, 
  citecolor   = blue 
}
\usepackage[all]{hypcap} 

\slugcomment{}

\shorttitle{Deep L-band VLA observations of the cluster 1RXS\,J0603.3+4214 }
\shortauthors{Rajpurohit et al.}

\begin{document}


\title{ Deep VLA observations of the cluster 1RXS\,J0603.3+4214 in the frequency range 1-2\,GH\MakeLowercase{z} }
\email{kamlesh@tls-tautenburg.de}


\author{K. Rajpurohit\altaffilmark{1}$^{\star}$, M. Hoeft\altaffilmark{1}, R. J. van Weeren\altaffilmark{2}, L. Rudnick\altaffilmark{3}, H. J. A. R\"{o}ttgering\altaffilmark{4}, W. R. Forman\altaffilmark{2}, \\ M. Br\"{u}ggen\altaffilmark{5},  J. H. Croston{\altaffilmark{6}}, F. Andrade-Santos\altaffilmark{2}, W. A. Dawson\altaffilmark{7}, H. T. Intema\altaffilmark{4}, R. P. Kraft\altaffilmark{2},\\ C. Jones\altaffilmark{2}, M. James Jee\altaffilmark{8,9}
} 
\affil{$\sp{1}$Th\"{u}ringer Landessternwarte, Sternwarte 5, 07778 Tautenburg, Germany \\
$\sp{2}$Harvard Smithsonian Center for Astrophysics, 60 Garden Street Cambridge, MA 02138, USA\\
$\sp{3}$ Minnesota Institute for Astrophysics, University of Minnesota, 116 Church St. S.E., Minneapolis, MN 55455, USA\\
$\sp{4}$Leiden Observatory, Leiden University, P.O. Box 9513, NL-2300 RA Leiden, The Netherlands\\
$\sp{5}$Hamburger Sternwarte, Universit\"{a}t Hamburg, Gojenbergsweg 112, 21029 Hamburg, Germany\\
$\sp{6}$School of Physical Sciences, The Open University, Walton Hall, Milton Keynes, MK6 7AA, UK\\
$\sp{7}$Lawrence Livermore National Lab, 7000 East Avenue, Livermore, CA 94550, USA\\
$\sp{8}$Department of Astronomy and Center for Galaxy Evolution Research, Yonsei University, 50 Yonsei-ro, Seoul 03722, Korea\\
$\sp{9}$Department of Physics, University of California, Davis, One Shields Avenue, Davis, CA 95616, USA\\
}

 
\begin{abstract}
We report L-band VLA observations of 1RXS\,J0603.3+4214, a cluster that hosts a bright radio relic, known as the Toothbrush, and an elongated giant radio halo. These new observations allow us to study the surface brightness distribution down to one arcsec resolution with very high sensitivity. Our images provide an unprecedented detailed view of the Toothbrush, revealing enigmatic filamentary structures. To study the spectral index distribution, we complement our analysis with published LOFAR and GMRT observations. The bright `brush' of the Toothbrush shows a prominent narrow ridge to its north with a sharp outer edge. The spectral index at the ridge is in the range $-0.70\leq\alpha\leq-0.80$. We suggest that the ridge is caused by projection along the line of sight. With a simple toy model for the smallest region of the ridge, we conclude that the magnetic field is below $5\,\rm\mu G$ and varies significantly across the shock front. Our model indicates that the actual Mach number is higher than that obtained from the injection index and agrees well with the one derived from the overall spectrum, namely ${\cal M}=3.78^{\tiny{+0.3}}_{-0.2}$. The radio halo shows an average spectral index of $\alpha=-1.16\pm0.05$ and a slight gradient from north to south. The southernmost part of the halo is steeper and possibly related to a shock front. Excluding the southernmost part, the halo morphology agrees very well with the X-ray morphology. A power-law correlation is found between the radio and X-ray surface brightness. 
\end{abstract}



\section{Introduction} \label{sec:intro}
Diffuse radio emission, associated with the intracluster medium, has been observed in a growing number of galaxy clusters, seen as radio relics and radio halos \citep{Feretti1996,Feretti2012,Bruggen2012,Brunetti2014}. Both of these types of radio sources have a typical size of about $1$\,Mpc. They show a typical synchrotron spectrum\footnote{$S(\nu)\propto\nu^{{\alpha}}$ with spectral index $\alpha$}, a clear indication of the existence of cluster-wide magnetic fields and relativistic particles. Both relics and halos are associated with merging clusters \citep{Giacintucci2008,Cassano2010,Finoguenov2010,vanWeeren2011} suggesting that cluster mergers play a major role in their formation. 

Radio halos are found at the center of galaxy clusters, have a smooth regular shape morphology and are usually unpolarized. Most halos are found only in merging systems possessing X-ray substructure \citep{Liang2000,Cassano2010,Basu2012,Cassano2013}. The radio emission from the halo typically follows the X-ray emission from the thermal gas \citep{Govoni2001}, suggesting a direct connection between the thermal and non-thermal components of the intracluster medium (ICM). However, there are also a few systems in which the radio halo emission does not appear to follow the X-ray emission very well, e.g. in the Coma cluster \citep{Brown2011}, Abell\,3562 \citep{Giacintucci2005}, and MACS\,J0717.5+3745 \citep{Bonafede2012,vanWeeren2017b}.  

The cosmic ray electron (CRe) acceleration mechanism responsible for radio halos is still disputed. There are two main models that have been proposed to explain the origin of halos. The primary electron model suggests that relativistic populations of electrons are re-accelerated to higher energies by magneto-hydrodynamical turbulence induced during mergers \citep{Brunetti2001,Petrosian2001}. The secondary electron model proposes that the CRe are the secondary products of hadronic collisions between thermal ions and relativistic protons present in the ICM \citep{Dennison1980,Blasi1999,Dolag2000}. The secondary electron model predicts the existence of gamma rays as one of the products of hadronic collisions. However, the non-detection of gamma rays in clusters \citep{Jeltema2011,Brunetti2012,Ackermann2014,Ackermann2016} appears to challenge this model.

Radio relics are large elongated sources located at the periphery of the merging galaxy clusters. Most of them are strongly polarized \citep{Govoni2004,Ferrari2008}, indicating the presence of ordered magnetic fields. They often display irregular morphologies and are believed to be produced by diffusive shock acceleration (DSA) at the merger shocks \citep[e.g.][] {Drury1983,Ensslin1998,Roettiger1999,Hoeft2007,Kang2011}. One of the well-studied examples of radio relics is CIZA J2242.8+5301, the ``Sausage-relic'' \citep{vanWeeren2010,Stroe2013,Hoang2017}. 

There is strong evidence that relics trace shock waves occurring in the ICM during cluster merger events. Cosmological shocks are capable of accelerating electrons \citep{Ryu2003} to relativistic energies. These electrons, together with magnetic fields of ${\mu}$G-level \citep[][and references therein]{Carilli2002}, emit synchrotron radiation. In the standard scenario for radio relics, particle acceleration \citep{Ensslin1998} is described by the DSA mechanism. For most of the relics, the observed properties such as the gradual spectral index steepening towards the cluster center, the high degree of polarization with magnetic field lines parallel to the source extension, and the integrated spectrum with a power-law form can be well explained by the DSA model.  

For several radio relics the jump related to the shock in X-ray surface brightness and temperature has been searched for and identified \citep{Sarazin2013,Stroe2013,Shimwell2015,vanWeeren2016,Eckert2016}. For some relics the derived X-ray Mach numbers are low, posing a severe challenge for the standard scenario for relics \citep{Akamatsu2012,vanWeeren2016}. For weak shocks, namely ${\hfil \mathcal{M}}$$\leq 3$, DSA is inefficient in accelerating particles from the thermal pool to relativistic energies. To solve this low Mach number issue, several alternative models were proposed such as the shock re-acceleration model \citep{Kang2011,Kang2012,Pinzke2013,vanWeeren2017b,vanWeeren2017a}. There are also a few systems where the Mach numbers of the shock waves inferred from X-rays are higher than those inferred from radio observations \citep{Akamatsu2012,Botteon2016}.


\setlength{\tabcolsep}{17pt}
\begin{table*}
\caption{ VLA  L-band observations overview}
\label{Table 1}
\centering
 \begin{threeparttable} 
\begin{tabular}{l l l l l }
\hline\hline
 & A-configuration  & B-configuration  & C-configuration & D-configuration   \\ 
\hline
Observation dates & Dec 8, 2012 & Nov 30, 2013 & June 17, 2013  & Jan 28, 2013\\
Frequency range & 1-2\,GHz  & 1-2\,GHz & 1-2\,GHz & 1-2\,GHz \\
Integration time & 1\,s & 3\,s & 5\,s & 5\,s \\
On source time & 4+4\,hr &8\,hr & 6\,hr & 4\,hr \\
\hline
\end{tabular}
\begin{tablenotes}[flushleft]
  \footnotesize
     \vspace{0.05cm}
   \item\textbf{Notes.} For all configurations the number of spectral widows is 16, the number of channels per spectral window is 64 and the channel width is 1\,MHz. Full Stokes polarization information was recorded. 
    \end{tablenotes}
    \end{threeparttable} 
\label{Tabel:Tabel1}
\end{table*}


\setlength{\tabcolsep}{7.0pt}   
\begin{table*}[!htbp]
\caption{Image properties }
\centering
\label{Table 2}
 \begin{threeparttable} 
\begin{tabular}{l c l l c c c c c }
\hline\hline
name$^{\dagger}$ & configurations$^{\ddagger}$ & restoring beam & weighting & uv cut & uv-taper  & RMS noise & RMS noise  & RMS noise\\ 
& (VLA) &&&&&(VLA)&(GMRT)&(LOFAR) \\
\hline
IM1& AB&  $1\farcs07 \times 0\farcs97$& Briggs &none&none&  2 $\mu$Jy&& \\
IM2 & ABCD & $1\farcs96 \times 1\farcs52$& Briggs &none&none&  6 $\mu$Jy&& \\
IM3 & ABCD & $3\farcs5\times 3\farcs5$& Briggs &none&$4\arcsec$&  7 $\mu$Jy&& \\
IM4 & ABCD &$5\farcs0 \times 5\farcs0$& Briggs &none& 6$\arcsec$ & 6 $\mu$Jy &&\\
IM5 & ABCD &$5\farcs5 \times 5\farcs5$& uniform & none& 6$\arcsec$ &  7 $\mu$Jy&& \\
IM6 & ABCD &$5\farcs5 \times 5\farcs5$& uniform & $\geq0.2$\,k$\lambda$& 6$\arcsec$ &  7 $\mu$Jy&&144 $\mu$Jy \\
IM7 & ABCD &$5\farcs5 \times 5\farcs5$& uniform & $\geq0.9$\,k$\lambda$& 6$\arcsec$ &  8 $\mu$Jy&60 $\mu$Jy&170 $\mu$Jy \\
IM8 & ABCD &$6\farcs5 \times 6\farcs5$ & uniform& $\geq0.2$\,k$\lambda$&8$\arcsec$ &  8 $\mu$Jy &&155 $\mu$Jy\\
IM9 & ABCD &$7\farcs0 \times 7\farcs0$ & Briggs&none&8$\arcsec$&  8 $\mu$Jy &&\\
IM10 & ABCD& $10\arcsec \times10\arcsec$& Briggs &none& 10$\arcsec$ & 11 $\mu$Jy &&\\
IM11 & ABCD& $11\arcsec \times11\arcsec$& uniform &none& 10$\arcsec$ & 11 $\mu$Jy &&\\
IM12 & ABCD& $11\arcsec \times11\arcsec$& uniform &$\geq0.2$\,k$\lambda$& 10$\arcsec$ & 12 $\mu$Jy &&176 $\mu$Jy\\
IM13 & ABCD&$15\arcsec \times15 \arcsec$& Briggs &none&16$\arcsec$& 16 $\mu$Jy &&\\
IM14 & ABCD&$16\arcsec \times16\arcsec$& uniform &$\geq0.2$\,k$\lambda$&16$\arcsec$& 18 $\mu$Jy&& 208 $\mu$Jy\\
IM15 & ABCD&$25\arcsec \times25 \arcsec$& Briggs & none& 25$\arcsec$& 24 $\mu$Jy &&190 $\mu$Jy\\
IM16 & ABCD&$25\arcsec \times25 \arcsec$& uniform & none& 25$\arcsec$& 22 $\mu$Jy &&186 $\mu$Jy\\
\hline \\
\\
\end{tabular}
\begin{tablenotes}[flushleft]
  \footnotesize
   \vspace{-0.6cm}
   \item\textbf{Notes.} Imaging was always performed using multi-scale clean, $\tt{nterms}$=2 and $\tt{wprojplanes}$=500. For all images made with {\tt Briggs} weighting we used ${\tt robust}=0$;\,$^{_\dagger}$image name;\,$^{_\ddagger}$uv-data of configuration combined for imaging.
    \end{tablenotes}
    \end{threeparttable} 
\label{Tabel:Tabel2}
\end{table*}

In this work, we present the results of deep L-\,band VLA observations of the diffuse radio emission associated with the merging galaxy cluster 1RXS\,J0603.3+4214. We complement our analysis with published LOFAR, GMRT and Chandra observations to study the spectral properties of the radio emission from the cluster and its possible relation to the thermal X-ray emission. We attempt to explain the observed surface brightness profiles by a model using a log-normal magnetic field distribution.

Throughout this paper we assume a $\Lambda$CDM cosmology with $H_{\rm{ 0}}=71$ km s$^{-1}$\,Mpc$^{-1}$,  $\Omega_{\rm{ m}}=0.3$, and $\Omega_{\Lambda}=0.7$. At the cluster's redshift, $1\arcsec$ corresponds to a physical scale of 3.64\,kpc. All images are in the J2000 coordinate system.


\section{1RXS\,J0603.3+4214} \label{sec:toothbrush}
The galaxy cluster 1RXS\,J0603.3+4214, located at \,\,$z=0.225$, is known to host three radio relics and a giant elongated radio halo \cite{vanWeeren2012a}. The most prominent and noticeable radio feature is a large bright relic in the north. It has a peculiar linear morphology, extending to about 1.9\,Mpc, with three distinct components resembling the brush\,(B1) and handle (B2+B3) of a toothbrush. The handle of the Toothbrush is enigmatic because of its large and straight extent and its asymmetric position with respect to the cluster merger axis. \cite{Bruggen2012}, using a hydrodynamical N-body simulation, reproduced the elongated linear morphology of the Toothbrush. They showed that a triple merger between two equal mass clusters merging along the north-south axis, together with a third, less massive, cluster moving in from the southwest, may cause the peculiar shape of the Toothbrush. 

For the Toothbrush, \cite{vanWeeren2012a} found a relatively flat radio spectral index of $\alpha\approx$ $-0.6$ to $-0.7$ (between 610 and 325\,MHz) at the northern edge. However, recent radio observations indicate a spectral index of $\alpha^{1500}_{150}=-0.80$ \citep{vanWeeren2016}. According to standard DSA, this spectral index suggest a Mach number of ${\hfil \mathcal{M}}=2.8^{+0.5}_{-0.3}$. \cite{Stroe2016} presented the integrated spectrum of the Toothbrush from 150\,MHz to 30\,GHz and detected a spectral break at frequencies above about 2\,GHz. On the other hand, \cite{Kierdorf2016} found that the spectrum can be well-fitted by a single power law below 8.35\,GHz, suggesting that a break in the spectrum does not exist below 8.35\,GHz. \cite{Basu2016} studied the impact of the Sunyaev-Zeldovich effect on the observed synchrotron flux and found that the radio spectrum is affected above 10\,GHz. 
 
\cite{Ogrean2013} observed the cluster with XMM-Newton and found two distinct X-ray peaks and evidence of density and temperature discontinuities, indicating the presence of shocks. The weak-lensing study showed that the cluster is composed of a complicated dark matter substructures closely tracing the galaxy distribution \citep{Jee2016}. The cluster mass is dominated by two massive clumps with a 3:1 mass ratio. Recently, Chandra observation revealed the presence of a very weak shock of a Mach number of ${\hfil \mathcal{M}}\approx1.2$ at the other edge of the brush \citep{vanWeeren2016}. Clearly, the Mach number of the shock estimated from X-ray observations is lower than that inferred from the observed radio spectral index. 

\cite{vanWeeren2012a} also found that the cluster hosts a $\sim$2\,Mpc large giant radio halo. The spectral index map of the halo revealed a remarkably uniform spectral index distribution, namely $\alpha=-1.16$, with an intrinsic scatter of $\leq0.04$ \citep{vanWeeren2016}. The halo and the brush region of the Toothbrush seem to be connected by a region with a spectral index of $\alpha=-2.0$ after which the spectrum gradually flattens to $\alpha=-1.0$ and returns to being uniform in the halo. \cite{vanWeeren2016} suggested that the flattening of the spectral index is due to the re-acceleration of `aged' electrons downstream of the relic by turbulence, indicating a connection between the northern relic and the halo.


\begin{figure*}[!thbp]
\centering
\hspace{-4.1cm}	
     \includegraphics[width=17.5cm]{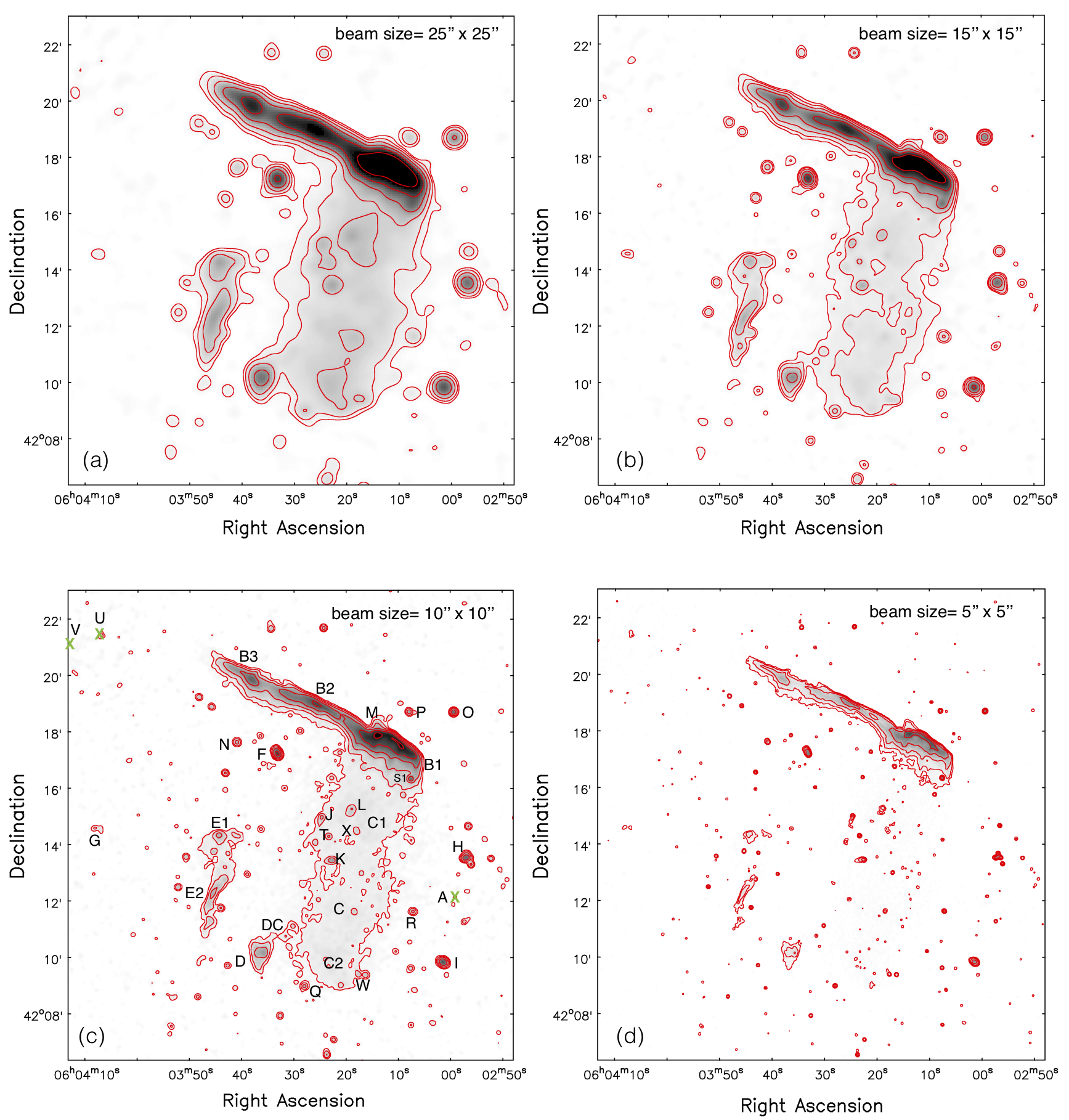}
 \hspace{-4.2cm}		
        \vspace{-0.16cm}
 \caption{ 
 	VLA 1-2\,GHz images of 1RXS\,J0603.3+4214 at different resolutions. In all four images the known relics (B, D and E) are evidently recovered. The image properties are given in Table\,\ref{Tabel:Tabel2}. Here, panel (a), (b), (c), and (d) correspond to IM15, IM13, IM10, and IM4, respectively. Contour levels are drawn at $[1, 2, 4, 8,\dots]\,\times4.5\,\sigma_{{\rm{ rms}}}$. In these images there are no region below $-4.5\,\sigma_{{\rm{ rms}}}$, except a small area close to the subtracted source A. The green cross marks the exact location of the sources which have been subtracted, namely A, U and V. } 
 \label{figure:Figure1}
\end{figure*}

 
\section{Observations and data reduction}
\label{sec:observations}
Observations of the cluster 1RXS\,J0603.3+4214 were made with the VLA  in L-band covering a wide frequency range of 1-2\,GHz (project code: SE0737, PI R. J. van Weeren). The L-band observations carried out with A, B, C, and D configurations are summarized in Table\,\ref{Tabel:Tabel1}. The VLA data correspond to a total integration time of around 26 hours on the target. A total bandwidth of 1\,GHz was recorded, spread over 16 spectral windows each divided into 64 channels. All four circular polarization products were recorded. For each configuration, the calibrator 3C147 was observed for around 30 minutes. The second calibrator 3C138 was observed at the end of the target observation for around 15 minutes. In order to cover the largest range of spatial scales and to maximize signal-to-noise, we combine all four VLA configurations datasets for imaging.

The data were calibrated and reduced with the $\tt{CASA}$\footnote{\url{http://casa.nrao.edu/}} package \citep{McMullin2007}, version 4.6.0. Initially, the data obtained from the four different configurations were separately calibrated. The first step of data reduction consisted of the Hanning smoothing of data. After this, we determined and applied elevation dependent gain tables and antenna offset positions. The data were then inspected for RFI (Radio Frequency Interference) removal. The ${\tt CASA}$ ${\tt tfcrop}$ mode was used for automatic flagging of strong narrow-band RFI. We then corrected for the bandpass using the calibrator 3C147. This prevents flagging of good data due to the bandpass roll-off at the edges of the spectral windows. The low amplitude RFI was searched for and flagged using ${\tt AOFlagger}$ \citep{Offringa2010}. The amount of data affected by RFI was typically only a few percent in A and B configurations but significant interference was encountered in C and D configurations.

As a next step in the calibration, we used the L-band 3C147 model provided by ${\tt CASA}$ software package and set the flux density scale according to \cite{Perley2013}. An initial phase calibration was performed using both the calibrators for a few neighboring channels per spectral window where the phase variations per channel were small. We corrected for the antenna delay and determined the bandpass response using the calibrator 3C147. After applying the bandpass and delay solutions, we proceeded to the gain calibration for both the calibrators. In the end, all relevant calibration solutions were applied to the target data. The resulting calibrated data were averaged by a factor of 4 in frequency per spectral windows and in time intervals of 10s, 10s, 6s, 4s in time for D, C, B, and A configurations, respectively. 

\begin{figure*}[!thbp]
  \begin{center}
   \vspace{0.2cm}
      \hspace{-1.2cm}
\includegraphics[width=18.0cm]{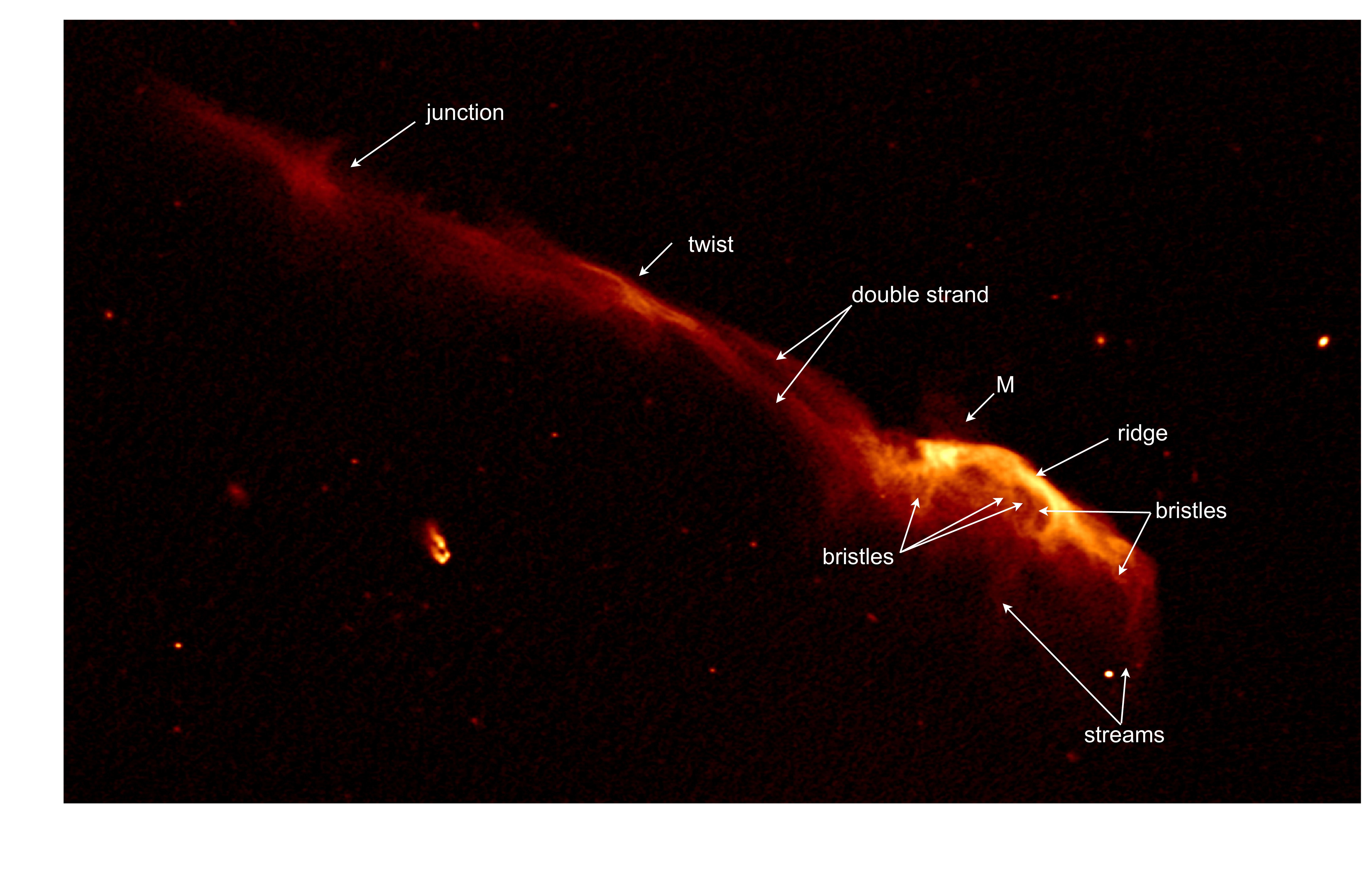}
         \hspace{-1.0cm}
 \vspace{-1.0cm}
  \caption{High resolution VLA 1-2 GHz image of the Toothbrush showing the complex, often filamentary structures. The image properties are given in Table\,\ref{Tabel:Tabel2}, IM2. }
\label{figure:Figure3} 
  \end{center}
\end{figure*}

After initial imaging of the target field, for each configuration, we carried out a self-calibration first with a few rounds of phase-only calibration followed by a final round of amplitude-phase calibration. For A and B configurations, we performed an additional bandpass calibration on the target, using the target field model derived from the self-calibration, which reduced the dynamic range. We visually inspected the self-calibration solutions and manually flagged some additional data. For wide-field imaging, we employed W-projection algorithm \citep{Cornwell2008} to correct for non-coplanarity. We imaged each configuration using ${\tt nterms}=3$ \citep{Rau2011} to take into account the spectral behavior of the bright sources. The deconvolution was always performed with ${\tt CLEAN}$ masks generated in the ${\tt PyBDSF}$ \citep{Mohan2015}. We used the ${\tt Briggs}$ weighting scheme with a {\tt robust} parameter of 0. 

After deconvolving each configuration independently, we subtracted from the uv-data three bright sources (labeled as A, V, and U in Figure\,\ref{figure:Figure1}). Next, for each configuration, we made a single deep Stokes I continuum image using the full bandwidth with ${\tt nterms}=2$ \text{that reduced the noise level by about 30-40\%}. To speed up imaging, we subtracted in the uv-plane all sources outside a cluster region. 

Finally, the A, B, C, and D configurations data were combined together to make a single deep full bandwidth Stokes I image, using multi-scale clean \citep{Rau2011} and $\tt{nterms}=2$. An overview of the image properties is given in Table\,\ref{Tabel:Tabel2}. The output images were corrected for primary beam attenuation. 

We assume an absolute flux calibration uncertainty of 4$\%$ for the VLA L-band \citep{Perley2013}. For any flux density measurement we estimate the uncertainty via 
\begin{equation}
\Delta S =  \sqrt {(0.04S)^{2}+{N}_{{\rm{ beams}}}\ (\sigma_{{\rm{ rms}}})^{2}}
\end{equation}
where $S$ is the flux density, $\sigma_{{\rm{ rms}}}$ is the RMS noise and $N$$_{{\rm{ beams}}}$ is the number of beams. We list configuration, resolution, imaging parameters and achieved noise level in Table \,\ref{Tabel:Tabel2}. All the VLA images, unless otherwise noted, were made using data from all four configurations. 


\section{Results: VLA radio continuum images}
\label{sec:results}
We show in Figure\,\ref{figure:Figure1} the resulting 1-2\,GHz VLA radio continuum images of 1RXS\,J0603.3+4214 created with different uv-tapers. The known diffuse emission features, namely the bright Toothbrush, the two radio relics to the west, and the large elongated halo, are evidently recovered in our observation. The sources are labeled following \citet{vanWeeren2012a} and extending the list. The properties of compact and diffuse sources in the cluster are summarized in Tables\,\ref{Tabel:Tabel3} and \,\ref{Tabel:Tabel4}. 


\setlength{\tabcolsep}{15pt}
\begin{table}[!htbp]
\caption{Flux density of compact sources in the cluster region }
\centering
\label{Table 3}
 \begin{threeparttable} 
\begin{tabular}{l c l}
\hline\hline
Source &  $S_{1.5\,\rm{GHz}}$& type  \\ 
  &  mJy &   \\
\hline
A & $286.0\pm28.0$ & quasar\\
F&  $5.78\pm0.27$ &double-lobed  \\
G &  $4.44\pm0.32$ & double lobed  \\
H &  $2.52\pm0.23$ &  double-lobed  \\
I & $ 4.27\pm0.53$ & spiral-galaxy \\
J &  $0.58\pm0.17$ &  head-tail    \\
K &  $0.95\pm0.21$&head-tail \\
L &  $0.62\pm0.22$ &  \\
N&  $0.65\pm0.12$ &spiral-galaxy \\
O &  $1.72\pm0.20$ &spiral-galaxy \\
P &  $0.58\pm0.13$ &\\
Q & $ 0.43\pm0.09$  &\\ 
R&  $0.38\pm0.08$ &\\
S1 &  $1.46\pm0.09$ &\\
T &  $0.33\pm0.07$&\\
U &  $24.80\pm2.10$ & double-lobed\\
V &  $65.60\pm6.01$ &  double-lobed \\
W &  $0.30\pm0.07$& head-tail \\
X &  $0.25\pm0.06$ &  \\
\hline
\end{tabular}
\begin{tablenotes}[flushleft]
  \footnotesize
   \item\textbf{Notes.} Flux densities are measured in the image IM4, see Table\,\ref{Tabel:Tabel2}. The type is determined from the radio morphology and the presence of an optical counterpart.
    \end{tablenotes}
    \end{threeparttable} 
\label{Tabel:Tabel3}
\end{table}


\begin{figure*}[!thbp]
\vspace{-0.6cm}
  \begin{center}
  \includegraphics[width=17.0cm]{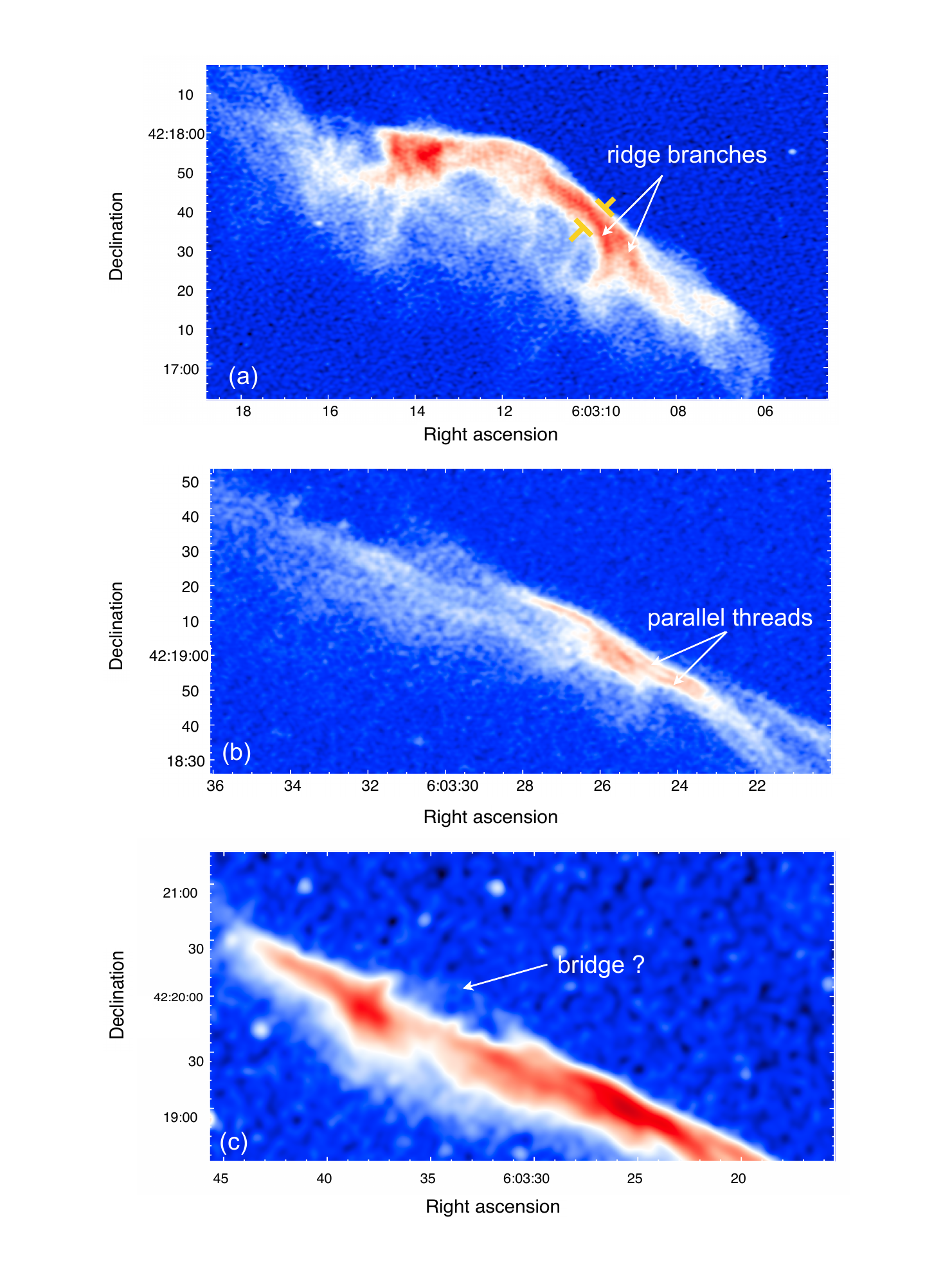}
\vspace{-1.0cm}
  \caption{VLA 1-2\,GHz images of different regions of the Toothbrush. {\it Top:} Cut-out around the B1 region, showing that the ridge consists of two parts which branch to the west. The ridge has a width of about 25\,kpc, measured across the region marked with yellow. The image properties are given in Table\,\ref{Tabel:Tabel2}, IM1. {\it Middle:} Cut-out around the center of B2 region (twist), revealing two very thin parallel filaments that are separated by 5\,kpc. The image properties are given in Table\,\ref{Tabel:Tabel2}, IM2. {\it Bottom:} Bridge connecting the B3 region of relic to B2.  The image properties are given in Table\,\ref{Tabel:Tabel2}, IM4. }
    \end{center}
\label{figure:ridge:bridge}
\end{figure*}


\begin{figure}[!thbp]
  \centering
  \hspace{-2.0cm}
    \includegraphics[width=9.0cm]{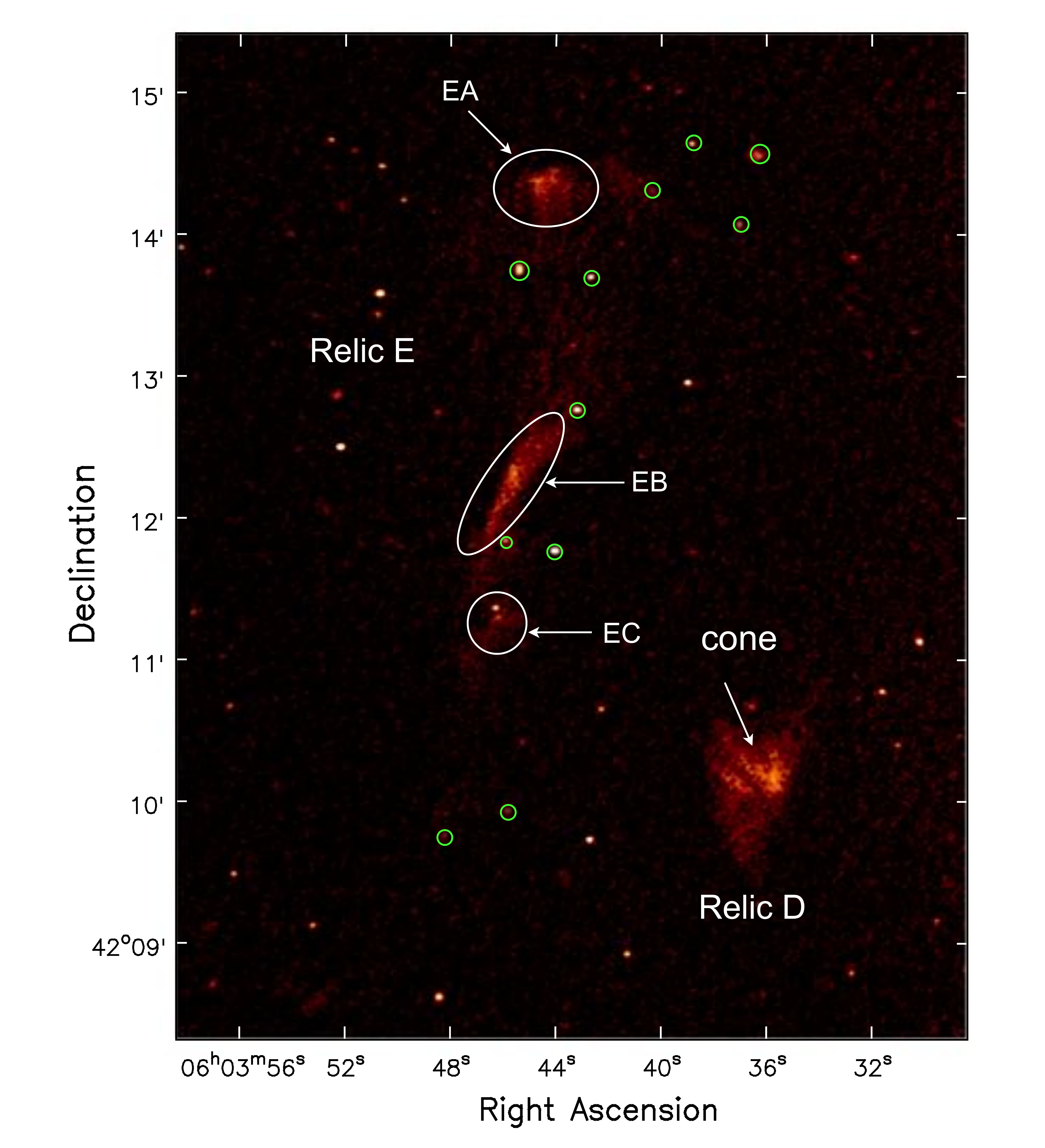}
      \hspace{-1.8cm}
      \vspace{-0.1cm}
 \caption{High resolution VLA 1-2\,GHz image of relics E and D. The image properties are given in Table\,\ref{Tabel:Tabel2}, IM2. The three bright compact regions EA, EB and EC are surrounded by diffuse low surface brightness emission. The point sources embedded within relic E, that are clearly detected at 1.5\,GHz with optical counterparts visible in the Subaru image, are marked with green circles.}
\label{figure:RelicED}
\end{figure}


\begin{figure*}[!thbp]
  \centering
  \hspace{-5.0cm}
    \includegraphics[width=18cm]{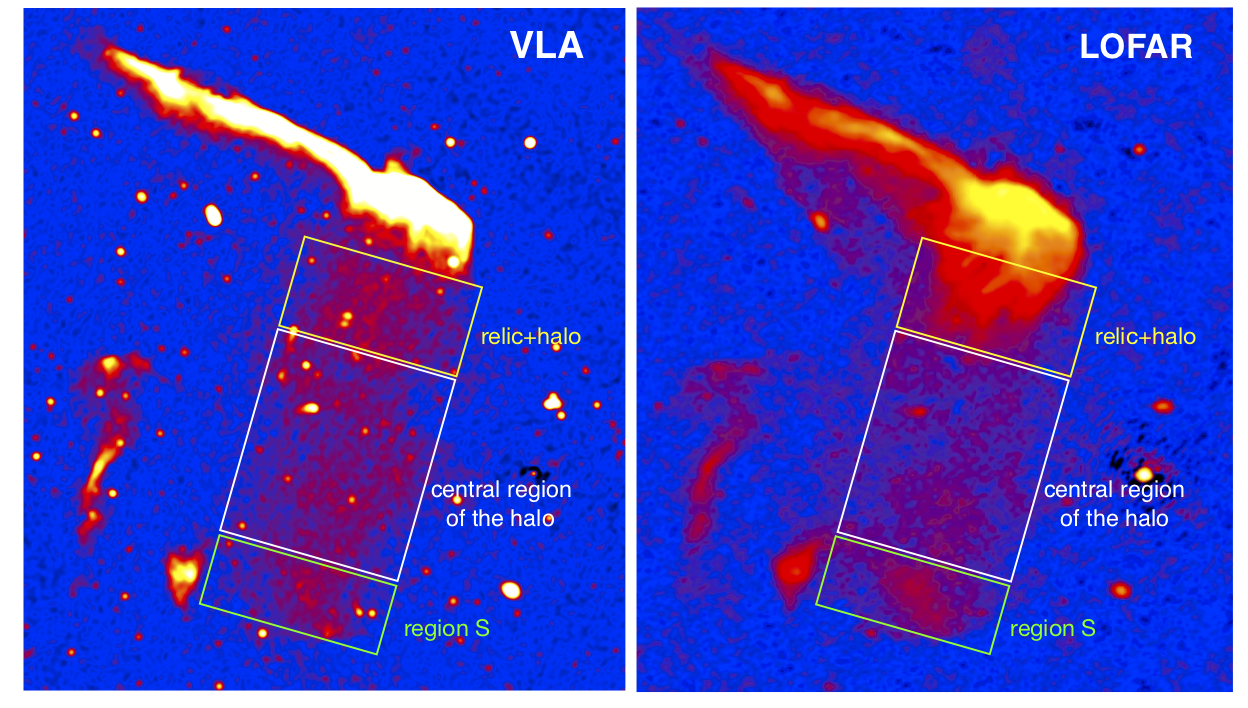}
      \hspace{-5.3cm}
            \vspace{-0.3cm}
 \caption{ Comparison of the halo between the VLA 1-2\,GHz and the LOFAR 120-181\,MHz image \citep{vanWeeren2016}. Both the images have similar resolution, the VLA image properties are given in Table\,\ref{Tabel:Tabel2}, IM9. To compare radio morphologies at these two frequencies, the colors in both images were scaled manually. In both images the yellow rectangular box indicates the `relic+halo' region (here, relic+halo indicates that in the VLA and LOFAR image the halo and relic surface brightness dominates, respectively. The southernmost part of the halo is denoted as region S. We define that region of the halo which excludes the relic+halo region and region S as `central region'. }
\label{figure:halocomparison}
\end{figure*}


\begin{figure*}[!thbp]
\begin{center}
\vspace{-0.35cm}
\includegraphics[width=17.8cm]{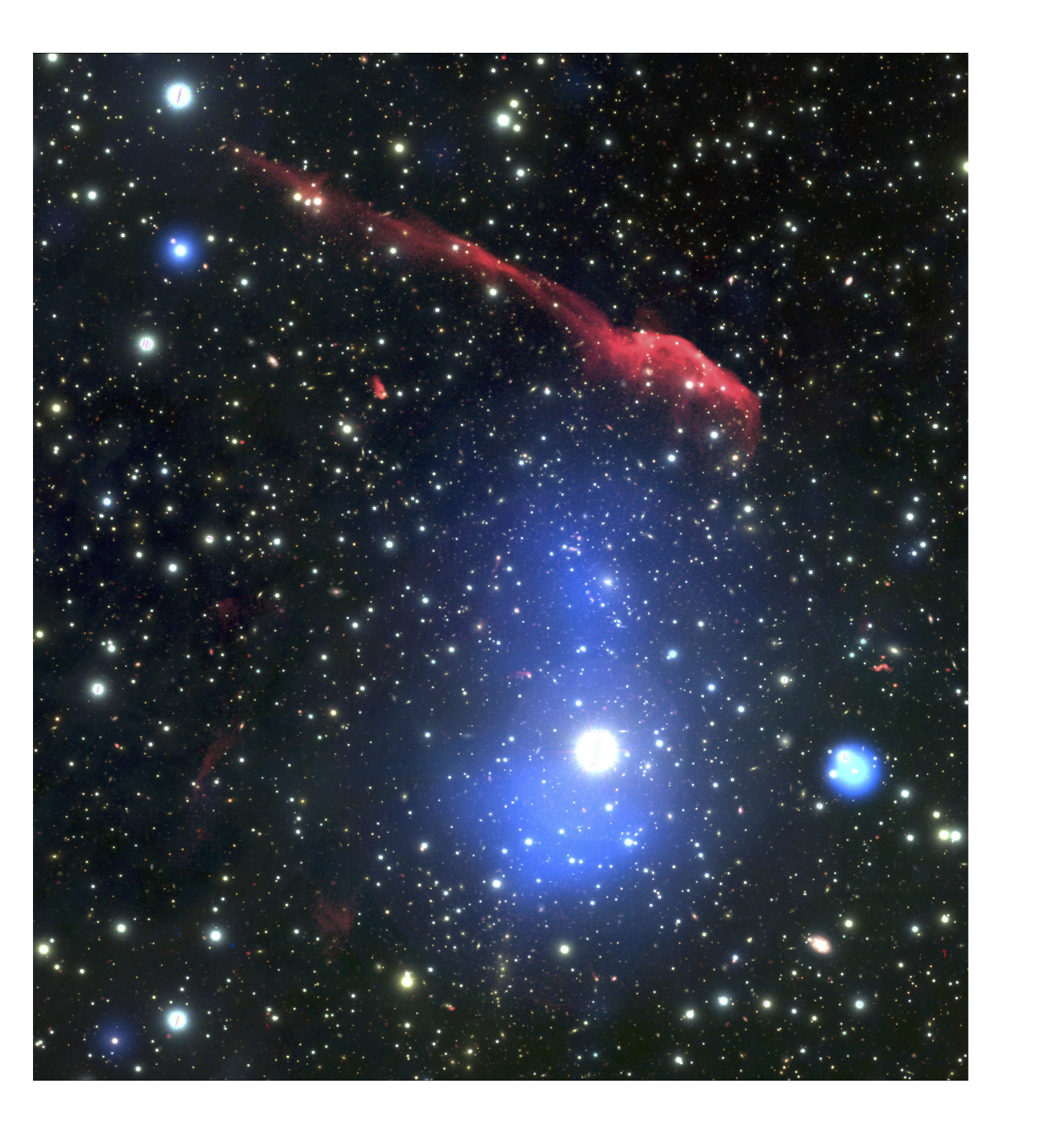}
\hspace{-1.0cm}
\vspace{-1.0cm}
 \caption{Radio, X-ray and optical overlay of cluster 1RXS\,J0603.3+4214. The intensity in red shows the radio emission observed with VLA  at a central frequency of 1.5\,GHz. The VLA image properties are given in Table\,\ref{Tabel:Tabel2}, IM3. The intensity in blue shows Chandra X -ray emission in the 0.5\,-\,2.0\,keV band and in the background is the color composite optical image created using Subaru data with g, r, and i intensities represented in blue, green, and red, respectively.}
\label{figure:overlay}
\end{center}
\end{figure*}


\subsection{Radio relics} 
\label{sec:relic}
A high-resolution VLA image of the Toothbrush with a restoring beam of $1\farcs96\times1\farcs52$ is shown in Figure\,\ref{figure:Figure3}. The complex, often filamentary structures in the Toothbrush at this resolution are one of the relic's most striking features. We briefly describe prominent features seen in Figure\,\ref{figure:Figure3}. The brush shows a distinct, more or less narrow ridge to the north with a quite sharp outer edge. At the narrowest location, the ridge has a width of about $25$\,kpc. 

To even better identify structures across the ridge, we further increased the resolution using only A and B configurations and achieving a restoring beam of $1\farcs07\times0\farcs97$, see Figure\,\ref{figure:ridge:bridge}. Interestingly, the ridge seems to consist of two parts which branch to the west, labeled as ridge branches in Figure\,\ref{figure:ridge:bridge} panel (a). The surface brightness downstream of the ridge drops very quickly in our image while at low frequencies (150\,MHz and 610\,MHz), the surface brightness decreases more gradually \citep{vanWeeren2012a,vanWeeren2016}.

The brush shows small filaments, `bristles', which are arc-shaped and more or less perpendicular to the ridge. The width of the bristles is about 3 to 5\,kpc. 
At 610\,MHz and $150$\,MHz \cite{vanWeeren2012a,vanWeeren2016} found several `streams' of emission, wider than the bristles, extending from the northern part of B1 to the south. These streams are also visible in the VLA 1-2\,GHz image, see Figure\,\ref{figure:Figure3}. To the north-east of B1, we confirm the low surface brightness emission, source M, reported by \citet{vanWeeren2012a}.

The origins of the filamentary structures, the ridge and the bristles, are not known. For Abell\,2256, \citet{Owen2014} reported many long pronounced filaments stretching across the entire relic which is presumably seen face-on. Possibly, the ridge and the bristles are caused by similar filaments. In contrast to the relic in Abell\,2256, the Toothbrush is seen edge-on, hence the bristles could be the ends of filaments stretched across the entire relic. 

Another distinctive morphological feature is the double strand, emerging from B1, see Figure\,\ref{figure:Figure3}. The intrinsic width of these strands varies from 30\,kpc to 17\,kpc when moving from west to east. The two strands merge at $06^{{\rm{h}}}03^{{\rm{m}}}25^{{\rm{s}}}$\, +$42^{\circ}$18${\arcmin}59{\arcsec}$ and further to the east there seem to be several strands again. It appears as if the strands were twisted (Figure\,\ref{figure:Figure3}). 

Interestingly, the highest resolution image reveals that in the twist region one of the strands actually consists of two thin parallel threads with a separation of about $1\farcs3$ corresponding to 5\,kpc, shown in Figure\,\ref{figure:ridge:bridge} panel (b).

The B3 part of the handle seems to consist of one filament running from east to west and one which runs arc-shaped from north to south. The highest surface brightness is found where the filaments cross. We denote this region as the `junction'\,(Figure\,\ref{figure:Figure3}). With a surface brightness close to the noise, a structure is tentatively visible which connects the northern tip of the junction and one end of the strands, therefore denoted as `bridge', see Figure\,\ref{figure:ridge:bridge} panel (c).  

The radio relic E, located on the eastern side of the cluster center is shown in Figure\,\ref{figure:Figure1}. It consists of two parts, E1 and E2. The total extent of E, from north to south, is about $5\farcm2$ corresponding to a physical size of 1.1\,Mpc. 

The high resolution image of relic E is shown in Figure\,\ref{figure:RelicED}. The three bright regions of relic E, labeled as EA, EB, and EC, are surrounded by low surface brightness emission. For none of the three regions can we identify a related radio galaxy. However, there are several radio galaxies embedded within relic E, marked with the green circles in Figure\,\ref{figure:RelicED}, but they don't appear to be connected with these three bright regions. This underlines that all three regions are diffuse emission.

Another diffuse emission region has been classified as a relic, namely source D, located south-west of relic E. We recover a similar morphology as found by \citet{vanWeeren2012a} using GMRT 610\,MHz data. Part of the emission resembles a bullet with a Mach cone, see Figure\,\ref{figure:RelicED}. Moreover, relic D is apparently connected to the halo via patchy emission DC, see Figure\,\ref{figure:Figure1}.  

Recently, evidence has been found that radio relics may originate from a combination of outflows of radio galaxies and the impact of a merger shock \citep{vanWeeren2017a}. Relics E and D show a quite unusual morphology, however, we do not find any connection to a nearby radio galaxy.


\subsection{Radio halo} 
\label{sec:halo}
The VLA images also recover the extended radio halo emission, shown in Figure\,\ref{figure:Figure1}. The halo emission is not detected at very high resolution, for example, Figure\,\ref{figure:Figure1} panel (d), due to its low surface brightness. However, we can use the sensitive high resolution image to identify faint compact sources in the halo region which may contribute to the total flux density measured at the lower resolution.

The shape of the halo in the VLA images is similar to the low frequency images \citep{vanWeeren2012a,vanWeeren2016}, elongated along the merger direction. At 1-2\,GHz the radio halo has a largest angular size of about $8\farcm7$ corresponding to a physical extent of 1.7\,Mpc. 

In the VLA image there is a slight decrease in the surface brightness, apparently separating the relic and the halo. It is also worth noting that the transition from halo to relic, i.e. which type of diffuse emission dominates the surface brightness, appears in the VLA image at a different location than in the LOFAR image, see Figure\,\ref{figure:halocomparison}.

In our sensitive high resolution radio maps, we detected 32 discrete sources ($>$$\,5\,\sigma_{{\rm{rms}}}$) within the halo region, including several head-tail radio galaxies. These sources were not detected in perviously published observations. Without subtracting the 32 sources, the measured flux density of the halo region at 10$\arcsec$ resolution (IM10) is $53.0\pm4.0\,$mJy.  We measure the flux density for the entire halo C. The combined total flux density of the 32 discrete sources is $\sim13$\,mJy, which means $\sim25\%$ of  total flux resides in these sources. The flux density of the halo hence amounts to $S^{\rm halo}_{{1.5\,\rm GHz}}\,\sim40.0$\,mJy. We also confirm that the southern part of the radio halo shows a sharp outer edge as reported by \cite{vanWeeren2016}.


\subsection{Optical, X-ray and radio continuum overlay}
\label{sec:overlay}
The cluster 1RXS\,J0603.3+4214 was observed with the 8.2\,m Subaru telescope on 25 February 2013 in r, g, and i colors for 2880\,s,  720\,s, and 720\,s, respectively \citep{Jee2016}. The X-ray emission of the cluster was observed with ACIS-I Chandra telescope (210 ks) in 2013 in 0.5-2.0\,keV band \citep{vanWeeren2016}. The spectroscopic redshifts of the sources in the field were taken from Subaru observations (Dawson et al. in preparation).

We create an overlay of the optical, X-ray, and radio emission in the cluster region using a Subaru composite image, the Chandra X-ray image, and our VLA 1-2\,GHz radio continuum image, see Figure\,\ref{figure:overlay}. To see the radio emission nicely, in particular the filamentary structures associated with the Toothbrush, we convolve the VLA image $1\farcs96\times1\farcs52$ image to $3\farcs5\times3\farcs5$.

For relic E, point sources for which we found an optical counterpart are denoted with green circles in Figure\,\ref{figure:RelicED}. We do not find an optical counterpart for the brightest regions of relic E, i.e. for EA, EB, and EC, which could be assumed to be the source of the radio emission in these regions.


\setlength{\tabcolsep}{4.5pt}
\begin{table*}[!htbp]
\caption{ Flux densities and integrated spectra of the diffuse radio sources in the cluster 1RXS\,J0603.3+4214 }
\centering
\begin{threeparttable} 
\begin{tabular}{ c | c | c | c | c | c | c | c | c | c}
\hline \hline
\multirow{1}{*}{Source} & \multicolumn{3}{c|}{VLA}  & \multirow{1}{*}{GMRT} & \multicolumn{2}{c|}{LOFAR} &\multirow{1}{*}{LLS} & \multirow{1}{*}{$\alpha^{1500}_{150}$} & \multirow{1}{*}{$\cal{M}$}\\ 
 \cline{2-4} \cline{4-5}\cline{6-7}
  &$25\arcsec$& $5\farcs5$ and $11\arcsec$ & $5\farcs5$ and $11\arcsec$ &$5\farcs5$& $25\arcsec$& $5\farcs5$ and $11\arcsec$& & &\\
   && & with uv cut  & with uv cut  & & with uv cut  & & &\\
  
   \cline{2-4} \cline{4-5}\cline{6-7}
  &$S_{{\rm{1.5\,GHz}}}$ & $S_{{\rm{1.5\,GHz}}}$& $S_{{\rm{1.5\,GHz}}}$ &  $S_{{\rm{610\,MHz}}}$ & $S_{{\rm{150\,MHz}}}$ & $S_{{\rm{150\,MHz}}}$ & & &\\
 &mJy & mJy & mJy & mJy & mJy & mJy &Mpc & &\\
  \cline{2-4} \cline{4-5}\cline{6-7}
  \hline  
 (1) & (2)  &(3)  & (4)  & (5) &(6) & (7) & (8) & (9)& (10)  \\

relic\,B& $310.0\pm21$ & $296.0\pm17.0$ & $258.0\pm14.0$ & $751.0\pm78.0$ & $4428.0\pm410.0$&$3669.0\pm378.0$ &1.9 & $-1.15\pm0.02$ & $3.78^{+0.3}_{-0.2}$ \\
halo\,C &$33.4\pm2.7$ & $31.6\pm2.6$ & $30.0\pm1.2$ &$-$& $490.0\pm56.0$& $441.0\pm44.0$ &1.7$^{\dagger}$ &$-1.17\pm0.04$ & $-$\\ 
relic\,D & $5.2\pm0.8$ & $4.9\pm0.7$ & $4.6\pm0.2$ &$13.0\pm1.7$ & $98.1\pm11.8$& $87.1\pm9.1$ & 0.3 &$-1.28\pm0.05$& $-$ \\ 
relic\,E & $11.6\pm1.3$ &$10.1\pm1.2$ & $9.0\pm0.3$ & $18.7\pm2.2$ &$153.1\pm13.0$& $115.2\pm11.8$ &1.1 &$-1.11\pm0.05$ & $4.3^{+1.4}_{-0.7}$ \\
region\,S &$9.1\pm1.1$ &$8.4\pm1.0$ & $7.7\pm0.3$ &$-$& $172.0\pm18.8$ &$148.0\pm15.0$ &0.7&$-1.28\pm0.05$ & $-$\\
\hline 
\end{tabular}
\begin{tablenotes}[flushleft]
\footnotesize
\item{\textbf{Notes.}} The regions where the fluxes were extracted are indicated in Figure\,\ref{figure:integrated} right panel. Column (1) Source name; Column (2) gives flux densities measured from the IM16, see Table\,\ref{Tabel:Tabel2} for image properties; Column (3) gives flux densities measured in IM5 and IM11 for relics and halo/region S, respectively; Column (4) gives flux densities measured in IM7 and IM12 for relics and halo/region S; Column (5) gives flux densities measured from the GMRT image with properties given in Column (4); Column (6) gives flux densities measured from the LOFAR image with properties given in Column (2); Column (7) gives flux densities measured from the LOFAR image with properties given in Column (4); Column (8) Largest linear size at 1.5\,GHz; Column (9) Integrated spectral indices between 150\,MHz and 1.5\,GHz obtained by fit to Column (4), Column (5) and Column (7). We assume an absolute flux scale uncertainty of 10\% for the GMRT and LOFAR data, and 4\% for the VLA data; Column (10) Mach numbers derived from the integrated spectrum given in Column (9);$^{_\dagger}$size of the entire halo C.
\end{tablenotes}
\end{threeparttable} 
\label{Tabel:Tabel4}   
\end{table*}


\section{Analysis and Discussion}
\label{sec:analysisdiscussion}
To study the spectral characteristics of the cluster radio emission, we use the VLA  1-2 \,GHz, the GMRT 610\,MHz and the LOFAR 120-181\,MHz observations. The GMRT and LOFAR data sets were originally published by \cite{vanWeeren2012a,vanWeeren2016}. We also use Chandra observations to study the X-ray emission from 1RXS\,J0603.3+4214. For data reduction steps, we refer to \cite{vanWeeren2012a,vanWeeren2016}. 

The radio observations reported here were performed using three different interferometers each of which has different uv-coverages. This results in a bias in the total flux density measurements, the integrated spectra and the spectra index maps. To overcome these biases, we create images at 150\,MHz, 610\,MHz and 1.5\,GHz with a common minimum uv cut and uniform weighting.  Imaging is always performed with multi-scale clean and ${\tt nterms}=2$. To reveal spectral properties of different spatial scales, we tapered images accordingly. The imaging parameters and the image properties are summarized in Table\,\ref{Tabel:Tabel2}.


\begin{figure*}[!thbp]

\begin{center}
\hspace{-2cm}
    \includegraphics[width=17.5cm]{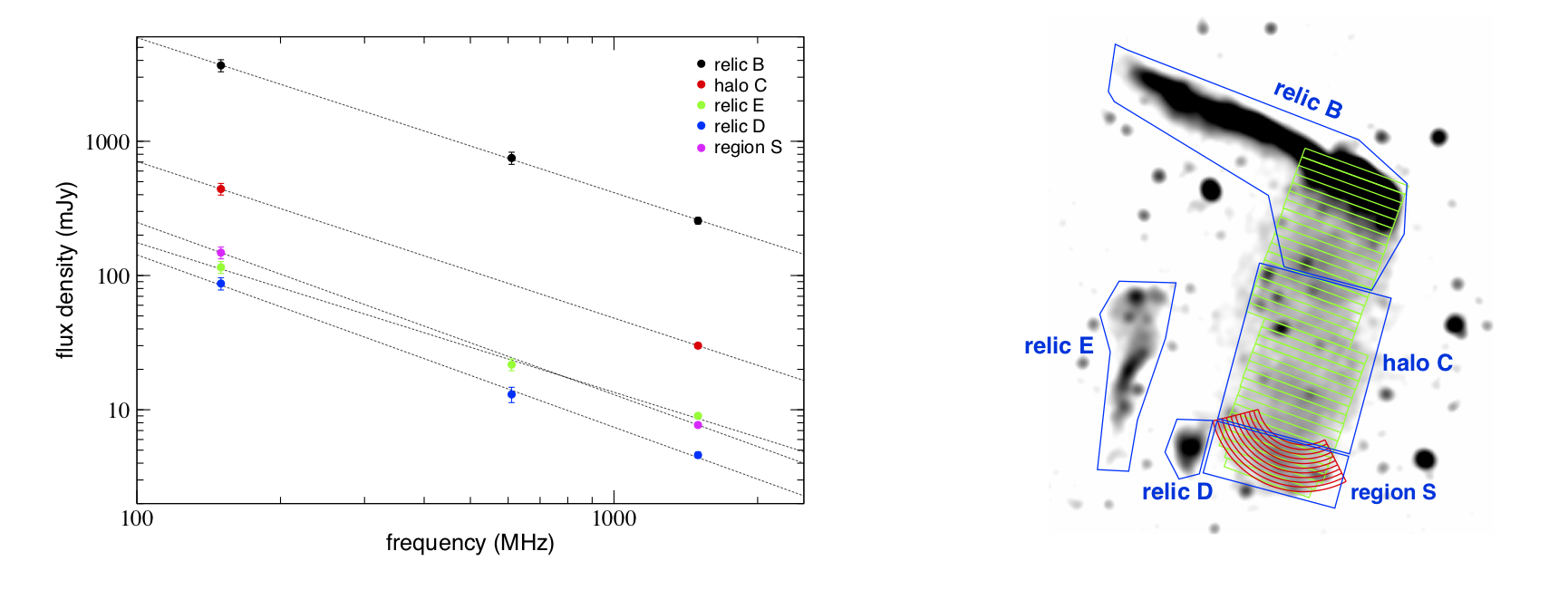}
    \hspace{-1.5cm}
\vspace{-0.4cm}
 \caption{{\it Left:} Integrated spectra of the radio sources B, C, E, and D and region S for 150\,MHz, 610\,MHz and 1.5\,GHz. Dashed lines are fitted straight power-laws with indices given in Table\,\ref{Tabel:Tabel4}. The radio spectrum of relics B, D, and E are well described by a single power law spectrum. The radio halo and region S is not detected in the GMRT 610\,MHz image. {\it Right:} VLA $16\arcsec$ resolution image depicting the regions where the integrated flux densities were measured. The green boxes have a width of $16\arcsec$ and were used to extract  the spectral index across the halo.}
 \end{center}
\label{figure:integrated}
\end{figure*}

\subsection{Flux measurements and integrated radio spectra}
\label{sec:int_spectrum}

It is difficult to accurately measure the flux density of extended low surface brightness sources in interferometric data. There are several reasons: \text{first} of all, if short baselines are missing or cut in the uv-data the flux of extended structures gets `resolved out'. Also, the distribution of antenna and the resulting uv-coverage may affect the flux measurement. Finally, structures with a surface brightness close to the noise level are notoriously difficult to deconvolve, the resulting flux measurements are very uncertain. Therefore, the deconvolution becomes particularly challenging for high resolution images. 


We investigate here how the measured flux densities changes when using different imaging parameters, for example, uv cut and resolution. In Table\,\ref{Tabel:Tabel4}, we report the flux measurement of several regions.  The flux densities reported in this section, unless stated otherwise, are measured from radio maps imaged with uniform weighting \citep{Stroe2016}. The VLA low resolution image made with all configurations and without any uv cut gives the maximum flux values. For instance, for relic B the flux density measured from the $25\arcsec$ resolution map is $310\pm21\,\rm mJy$. The flux density of the same region when measured from the $5\farcs5$ resolution image is  $296\pm17\,\rm mJy$. This means that the increase in the image resolution results in a 5\% flux loss. However, the image made using the same resolution but with a common inner uv cut of $0.9\,\rm k\lambda$ give a flux density of $258\pm14\,\rm mJy$. Hence, the uv cut causes an additional flux loss of about 15\%. In total, 20\% of the flux density reduction is caused by the resolution and the uv cut. For the LOFAR and GMRT data, we notice the same flux density reductions, e.g. without any uv cut the flux density of relic B at 150\,MHz is $4428\pm410\,\rm mJy$ and with an uv cut it is $3669\pm378\,\rm mJy$. Therefore, to ensure that we recover flux on the same spatial scales, we produce images with a common lower uv cut of $0.9\,\rm k\lambda$. Here, $0.9\,\rm k\lambda$ is the minimum uv-distance of the GMRT data. We also measure the flux densities from the 40\arcsec maps. The obtained flux densities are only marginally higher than the values reported in Table\,\ref{Tabel:Tabel4} at 25\arcsec resolution. To obtain the integrated spectrum of sources, we prefer high resolution imaging where it is easier to separate the source emission from other complex sources. The low resolution is not preferred because it is hard to decide the regions where to measure the flux density. We note that for all images multi-scale clean has been used. Although we lose some flux, less than 5\% as mentioned above, when measuring the flux density from higher resolution images, this effect remains same for all data sets and will not affect the integrated spectral index calculations.

According to the diffusive shock acceleration (DSA) theory, the particle acceleration at the shock front is determined by the Mach number of the shock \citep{Blandford1987}. More precisely, DSA in the test-particle regime generates a population of relativistic electrons with a power-law energy distribution $\partial N/ \partial E \propto E^{-\delta_{\rm inj}}$. The resulting radio spectrum is a power-law with index $\alpha_{\rm inj} = -(\delta_{\rm inj}-1)/2$ and reflects the Mach number of the shock front 

\begin{equation}
  \mathcal{M} 
  = 
  \sqrt{
       \frac{2\alpha_{\rm inj}-3}
            {2\alpha_{\rm inj}+1}  }  \: 
 \end{equation}
where $\alpha_{\rm inj}$ is the injection spectral index. For a stationary shock in the ICM, i.e. the electron cooling time is much shorter than the time scale on which the shock strengths or geometry changes, the integrated spectrum, $\alpha_{\rm int}$, is by 0.5 steeper than the injection spectrum, 
\begin{equation}
\alpha_{\rm int}  = \alpha_{\rm inj} + 0.5
 \end{equation}
Recent simulations by \cite{Kang2015} suggest that merger shocks might not be considered as stationary. The resulting spectra would be curved. We investigate here the integrated spectra of the relics and the halo.

The radio spectrum of the Toothbrush has been studied extensively. \citet{vanWeeren2012a} reported that the integrated spectrum from 74\,MHz to 4.9\,GHz has a power-law shape with $\alpha_{\rm int}=-1.10\pm 0.02$. \citet{Stroe2016} studied the integrated spectrum from 150\,MHz to 30\,GHz, using a uv cut of $0.8\,\rm k\lambda$, and detected a spectral steepening above about 2\,GHz. They reported that the spectral index steepens from $\,\alpha=-1.00$ to $\,\alpha=-1.45$. Recently, \citet{Kierdorf2016} studied the Toothbrush at 4.85\,GHz and 8.35\,GHz with the 100m-Effelsberg telescope and found a power-law spectrum with index $\alpha=-1.0\pm0.04$ up to 8.35\,GHz. Hence, it remains unclear if the integrated spectrum of the Toothbrush is curved at high frequencies. 

To obtain the integrated spectra of relics, we convolved the LOFAR, GMRT and VLA images to the same beam size of $5\farcs5\times5\farcs5$. The regions where the fluxes were extracted are indicated in Figure\,\ref{figure:integrated} right panel. We note that there is a region which belongs to the relic according to the surface brightness at 150\,MHz but at 1.5\,GHz to the halo. This region is denoted as `relic+halo' in Figure\,\ref{figure:halocomparison}. When determining the integrated spectrum, this region is considered as the part of relic B because in the VLA high resolution images, e.g. in Figure\,\ref{figure:Figure1} panel (d), this region does not contribute much to the total flux and will not significantly affect our measurements. We will argue in Section\,\ref{sec:haloanalysis1} that the southern part of the halo, denoted as `region S' in Figure\,\ref{figure:halocomparison}, may have a different origin. Hence, we exclude this as well as the `relic+halo' region when computing the halo spectrum. 

The radio halo and region S are not detected in the high resolution GMRT 610\,MHz image. Therefore, to obtain the integrated spectra of the radio halo and region S, we create new radio maps at 150\,MHz and 1.5\,GHz using a common lower uv cut of $0.2\,\rm k\lambda$ to match the scale of the VLA with LOFAR. We then convolve the LOFAR and VLA images to the same beam size of $11\arcsec\times11\arcsec$. 

The integrated synchrotron spectra for relics B, D, and E obtained by combining our measured flux densities at frequencies 150\,MHz, 610\,MHz, and 1.5 \,GHz, are shown in Figure\,\ref{figure:integrated} left panel. For the Toothbrush, we find a spectral index of $-1.15\pm0.02$ which is consistent with the previous values of $-1.10\pm0.02$ \citep{vanWeeren2012a} and $-1.09\pm0.05$ \citep{vanWeeren2016}. The spectrum is fitted well by a single power law and we do not find any evidence for a spectral steepening in this frequency range. Using equations (2) and (3), we derive a Mach number of ${\cal M}=3.78^{+0.3}_{-0.2}$. 

For relic E, we measure an integrated spectral index of $-1.11\pm0.05$, yielding a Mach number of $4.3^{+1.4}_{-0.7}$. The integrated spectral index of relic D is $-1.28\pm0.05$.

For a few relics the injection spectral indices derived from the integrated spectral indices are even flatter than allowed by DSA test particle theory, i.e. spectra have been found flatter than $-0.5$, for example in A2256 \citep{vanWeeren2012b,Trasatti2015} and CIZA\,J2242.8+5301 \citep{Kierdorf2016,Hoang2017}. The derived integrated spectral index of relics B, D and E are consistent with the DSA approximation. 

A steepening of the halo spectrum with frequency is noticed in Abell\,3562 and Abell\,2256 \citep{Giacintucci2005,vanWeeren2012b}. The turbulent acceleration model with particles emitting in a region with relatively uniform magnetic field intensity was invoked to explain the high frequency steepening in Abell\,3562 while the low frequency steepening in Abell\,2256 was explained with inhomogeneous turbulence.  For the radio halo in 1RXS\,J0603.3+4214, we find an integrated spectral index of $-1.17\pm0.04$.  

The integrated spectral index of the southernmost part of the halo, labeled as `region S' in Figure\,\ref{figure:halocomparison}, is $-1.28\pm0.05$, indicating that this region is steeper than the rest of the halo.

It is worth to emphasizing that the radio spectrum of relics B, D, and E are well described by a single power law spectrum and the flux densities presented here have been measured from high resolution images that were created using the same uv cut and weighting scheme at all frequencies.


\begin{figure*}[!thbp]
\begin{center}
 \vspace{-0.3cm}
\includegraphics[width=15cm]{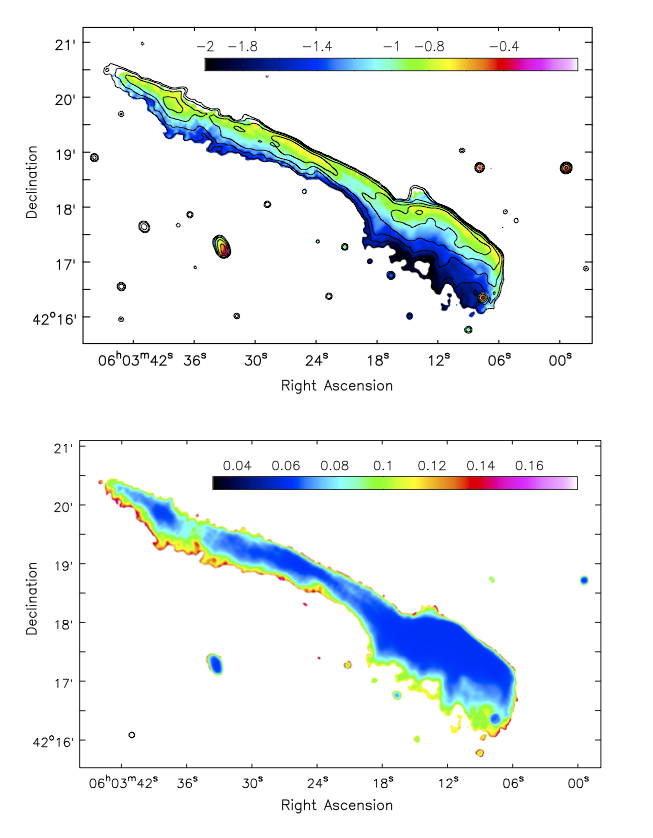}
\vspace{-0.7cm}
  \caption{{\it Top:} Spectral index map of the Toothbrush between 150\,MHz and 1.5\,GHz at $5\farcs5$ resolution, overlaid with the VLA contours. The image properties are given in Table\,\ref{Tabel:Tabel2}, IM6. Contours show the VLA flux density distribution. The contour levels are drawn at $[1, 2, 4, 8,\dots]\,\times\,4.5\,\sigma_{{\rm{ rms}}}$. The color bar shows spectral index $\alpha$ from $-0.2$ to $-0.4$. {\it Bottom:} Corresponding spectral index uncertainty map.}
  \end{center}
  \label{figure:Toothbrush_spectral_index}
\end{figure*}


\subsection{Analysis of the relics }
\label{sec:relics_spectra}
The Toothbrush is known to show a clear spectral index gradient \citep{vanWeeren2012a}. This is believed to reflect the aging of the relativistic electron population while the shock front propagates outwards \citep{vanWeerenSci}. However, it is worth noting, \cite{Skillman2013} found in simulations that relics can show a similar spectral index gradient, despite no inclusion of spatially resolved spectral aging in the model. In these simulations, the spectral index gradient is caused by a variation of the Mach number across the shock surface.

In \cite{vanWeeren2016}, a high resolution spectral index map of the Toothbrush was derived using the LOFAR 150\,MHz and GMRT 610\,MHz data. The latter restricted the spectral index map resolution to $6\farcs5$. Our VLA data allow us to reconstruct the surface brightness distribution at 1.5\,GHz with a higher resolution, hence we aim for high frequency spectral index maps, using LOFAR and VLA data, with the best resolution possible. 

We briefly describe how we created the spectral index maps. We apply an inner uv cut of 0.2\,$\rm{}k\lambda$ to the LOFAR data. This ensures that `resolving out' a possible large-scale flux distribution has a similar effect on both data sets. Here, 0.2\,$\rm{}k\lambda$  is the minimum uv-distance in the VLA data set. For imaging, we use uniform weighting and allow for spectral slopes (${\tt nterms}=2$). Due to different uv-coverages of the LOFAR and VLA data, the resulting images have slightly different resolution. Therefore, we convolve the VLA image to LOFAR resolution, i.e. $5\farcs5$. To have the same pixel position in both the images we use the CASA task {\tt imregrid}. Pixels with a flux density below 5$\,\sigma_{{\rm{ rms}}}$ in either image were blanked. Finally, we computed pixel-wise the spectral index maps.

 \begin{figure*}[!thbp]
\centering
 \hspace{-2.6cm}
\vspace{-0.31cm}
        \includegraphics[width=17cm]{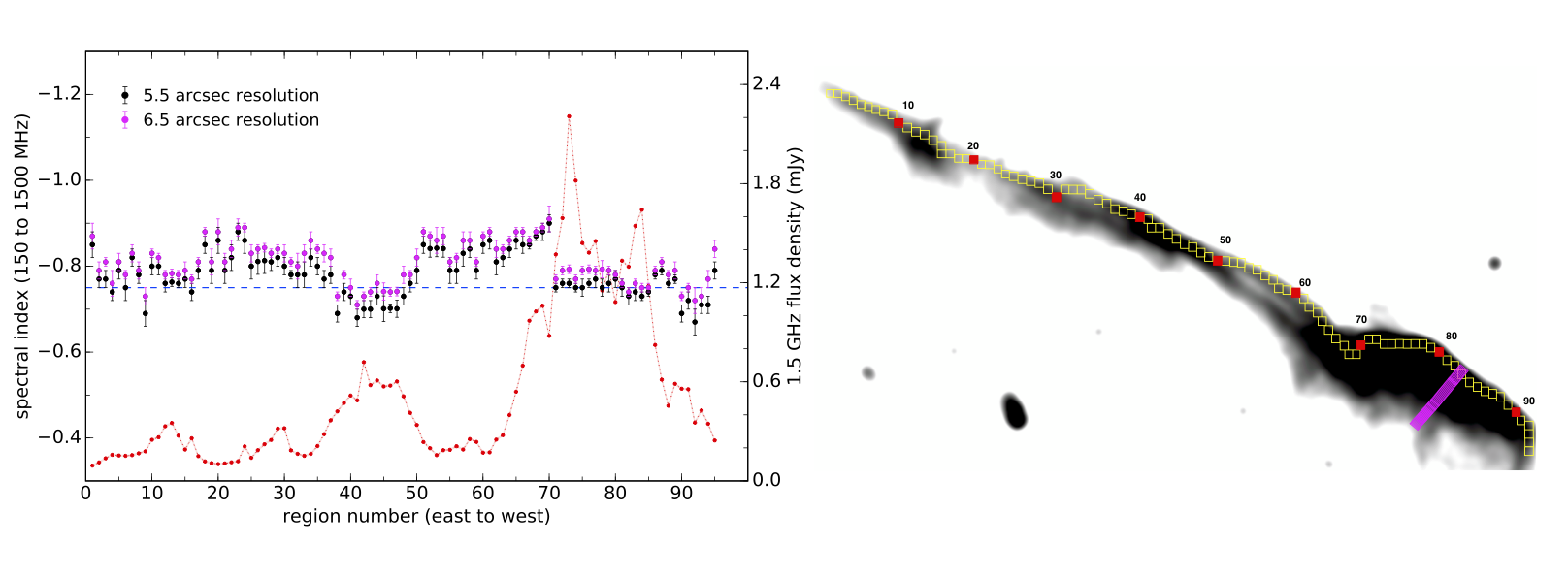}
 \hspace{-3.0cm}
 \vspace{-0.2cm}
  \caption{{\it Left:} Extracted spectral index between150 and 1500\,MHz across relic B1, from north to south, with the distance increasing from east to west. The black dots show the spectral index distribution at a resolution of $5\farcs5$ while the magenta at $6\farcs5$ resolution. The dashed blue horizontal line indicates the average spectral index of $\alpha=-0.75\pm0.05$ at the ridge. A shift in the spectral indices is clearly visible across the entire relic caused by lowering the resolution. The red dots trace the VLA 1.5\,GHz flux density along the relic, corresponding to the spectral indices shown with the magenta color, revealing a correlation between the brightness and the spectral index. Systematic uncertainties in the flux-scale were included in the error bars. {\it Right:} Box distribution across relic B overlaid on the VLA total intensity map at $5\farcs5$ resolution. The width of the boxes used to extract the indices is $5\farcs5$. The magenta coloured region is used to study the ridge. To obtain the flux densities, we create small rectangular boxes, inside the magenta region, with a width of $0\farcs7$ corresponding to 2.5 kpc. The total length of the magenta region is about 164 kpc.}
\label{figure:relicB_spectra}
\end{figure*}

To derive the spectral index uncertainty map, we take into account the image noise and the absolute flux calibration uncertainty. The flux scale uncertainty $f_{{\rm{err}}}$ of VLA L-band is about 4\,\% \citep{Perley2013} and for LOFAR we assume it to be 10\,\%. We estimate the spectral index error via:
\begin{equation}
\Delta{\alpha}  =  {\frac{1}{\ln\left( {\frac{\nu_1}{\nu_2}}\right)}} {\sqrt{{{\left({\frac{\Delta S_1}{S_1}}\right)}^2 + { { \left({\frac{\Delta S_2}{S_2}}\right)}^2}}}}
\end{equation}
where $S_1$ and $S_2$ are the flux density values at each pixel in the VLA and LOFAR maps at the frequencies $\nu_{1}=1500$\,MHz and $\nu_{2}=150$\,MHz. $\Delta S_1$ and $\Delta S_2$ were calculated as: 
\begin{equation}
\Delta S_{j} =  \sqrt {(f_{{{\rm{ err}}}} \times S_{j})^{2}+(\sigma^ {j}_{{{\rm{ rms}}}})^{2}} 
\end{equation}
where $j \in 1,2$ and $\sigma^{j}_{{{\rm{ rms}}}}$ is the corresponding RMS noise in the image. 

The $5\farcs5$ resolution spectral index map of the Toothbrush between 150\,MHz and 1.5\,GHz is shown in Figure\,\ref{figure:Toothbrush_spectral_index}. The map evidently confirms the remarkably uniform spectral steepening perpendicular to the relic extension, varying roughly from $-0.68$ to $-2.0$. These values are in agreement with \cite{vanWeeren2016}. The B2 region, where the double strand appears twisted, shows the flattest spectral index, namely $\alpha=-0.68\pm0.06$. The B3 region also shows few patches of flat spectral index.  

From our high resolution spectral index map, interesting details become visible: at the ridge the spectral index is apparently flatter than reported earlier \citep{vanWeeren2016}, namely $-0.70\leq\alpha\leq-0.80$. Interestingly, the spectrum already steepens across the ridge from $\alpha=-0.70$ to $-0.96$. In the double strand, it becomes evident that at the northern strand the spectral index is flatter and downstream of it, the spectrum is steeper. Surprisingly, at the southern strand the spectrum again gets flatter. The change in the spectral index across the double strand might be due to a new injection. Another possible explanation for the change of the spectral index across the double strand is an increase in strength of the local magnetic field which brightens up the emission and flattens the spectrum.We confirm the spectral index steepening along the `streams' of emission that emerges from the brush, reported by \cite{vanWeeren2016}. A clear north-south gradient is also seen across the streams, with steepening up to $\,-2.0\pm0.10$ at the southern ends. However, there is no evidence of the spectral index flattening at the southern end of the streams as found by \cite{vanWeeren2016}.

We investigate the spectral index distribution along the northern edge of the Toothbrush, with distance increasing from east to west. We determine the spectral indices in small square regions, see Figure\,\ref{figure:relicB_spectra} right panel, with sizes approximately equal to the restoring beam size of the map, i.e. $5\farcs5$. The extracted spectral indices are shown in the Figure\,\ref{figure:relicB_spectra} left panel with black dots. The spectral index varies mainly between $-$0.70 to $-$0.90, with brighter parts corresponding to flatter spectra. At the shock front, we find that the spectral index is roughly about $-0.75$, which is consistent with what we estimated from the spectral index maps. However, as mentioned above, these values are slightly flatter than reported by \cite{vanWeeren2016}. To compare our spectral index map with that of \cite{vanWeeren2016}, we also create a spectral index map at a $6\farcs5$ resolution. Such a comparison will also show the impact of the resolution on the spectral indices.

The magenta dots in Figure\,\ref{figure:relicB_spectra} left panel show the variation of the spectral index extracted from $6\farcs5$ resolution map. It is evident that the spectral index at the location of shock front, in this case, is  about $-0.80$ and is consistent with \cite{vanWeeren2016}. Evidently, there is a shift in the spectral indices across the entire radio relic caused by lowering the resolution. This raises the question how we can identify the actual injection spectrum at the shock front. The resolution study presented here indicates that the spectral index flattens with increasing resolution, hence with an even better resolution, we may find even flatter spectral indices. However, the origin of this resolution dependent flattening might be more complex, as we will report in Section\,\ref{sec:downstreamprofiles}.


\begin{figure*}[!thbp]
\begin{center}
            \hspace{-0.5cm}
    \includegraphics[width=17cm]{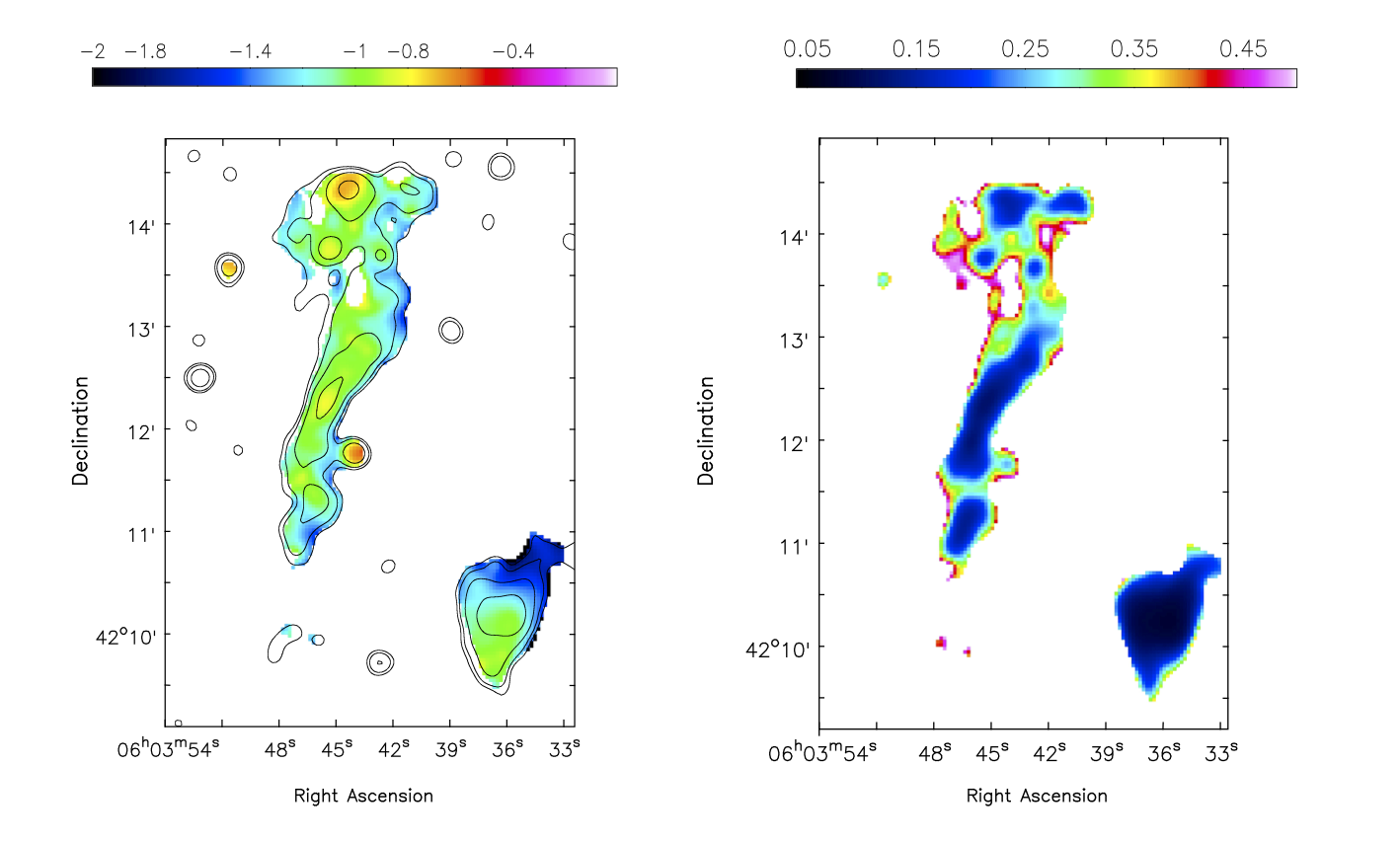}
                \hspace{-2.0cm}
            \vspace{-0.4cm}
  \caption{{\it Left:} Spectral index map for the relics E and D between 150 and 1500\,MHz at 11.0$\arcsec$ resolution. The image properties are given in Table\,\ref{Tabel:Tabel2}, IM12. Contours show the VLA flux density distribution. The contour levels are from the VLA and drawn at $[1, 2, 4, 8,\dots]\,\times4.5\,\sigma_{{\rm{ rms}}}$. The color bar shows spectral index $\alpha$ from -2.2 to -0.4. {\it Right:} Corresponding spectral index uncertainty map.} 
\end{center}
\label{figure:relicED_spectral_indexmap}
 \end{figure*}

The radio brightness distribution across the northern edge of the Toothbrush, with distance increasing from east to west, is displayed in Figure\,\ref{figure:relicB_spectra}. It is evident that the relic brightness is not uniform across its extension. 

The comparison of the spectral index and brightness distribution along the northern edge of the Toothbrush, from east to west (same region where we extracted spectral indices) reveals that there is a correlation between these two, i.e. brighter regions tends to be flatter. Variations on the order of 0.2 in the spectral index are seen in the brightest regions, like in the brush. For a curved electron energy distribution, an increase in the magnetic field strength will increase the emissivity which brightens the emission and flattens the spectrum \citep{Ellison1991}. We carried out a Spearman's rank correlation test to check the possible correlation between spectral index and brightness. From this test, we find that the p-value is less than 0.05, which indicate that correlation is statistically significant. 

The spectral index map for fainter relics is shown in Figure\,\ref{figure:relicED_spectral_indexmap}. To create the spectral index map for relic E and D, we convolve the VLA and LOFAR image to a common resolution of $11\arcsec \times11\arcsec$. 

The spectral indices in relic E range from $-0.70$ to $-1.50$. For some regions of relic E, our high resolution spectral index map reveals a relatively flat spectral index of about  $-0.70\pm0.10$, see Figure\,\ref{figure:relicED_spectral_indexmap}. We investigate regions showing flatter spectral index and searched for compact radio sources with optical counterparts. From the radio-optical overlay (Figure\,\ref{figure:overlay}), it is clear that the brightest regions (EA, EB and EC), showing a flat spectral index, namely $\alpha=-0.70$, do not show any optical counterparts.

Relics are expected to show a spectral index gradient, as the Toothbrush or Sausage-relic \citep{vanWeeren2010,vanWeeren2012a,vanWeeren2016,Hoang2017}. We do not find a significant spectral index gradient towards the cluster center, i.e. from east to west, for relic E.

For relic D we confirm the southwest spectral index steepening, varying from $-0.85$ to  $-$1.70,  as reported by \cite{vanWeeren2016}. They found the spectral index for relic D is in the range of  $-$0.90 to $-$1.40. It is interesting to see that spectrum flattens from south to west while the bright region of relic D (cone) is at the west which shows a relatively steeper spectrum.

Relics E and D show a quite unusual morphology, however, we do not find any connection to a nearby radio galaxy. Given the relatively small size and peculiar morphology of relic D, we speculate it could trace revised fossil radio plasma, possibly be re-accelerated by a merger induced shock. A similar scenario could also hold for relic E. It should be noted that due to the rather flat spectral index along the 1.1 Mpc extent of relic E, the source cannot simply be an old radio tail as spectral ageing would have resulted in a much steeper spectral index along the tail extent.


\subsection{The ridge} 
\label{sec:downstreamprofiles}

We can study the brush region with a resolution as good as about $1\arcsec$ due to its high surface brightness. Figure~\ref{figure:ridge:bridge} reveals its complex structure. A remarkable feature is the bright ridge with a relatively clear boundary. At the most narrow region the ridge has a width of about $25\,\rm kpc$. This raises the question about its origin. The high surface brightness across the ridge could be a projection effect of a shock front parallel to the line of sight. Alternatively, the ridge could originate from enhanced synchrotron emission due to a magnetic filament.

\begin{figure}
	\begin{center}
		\includegraphics[width=0.45\textwidth]{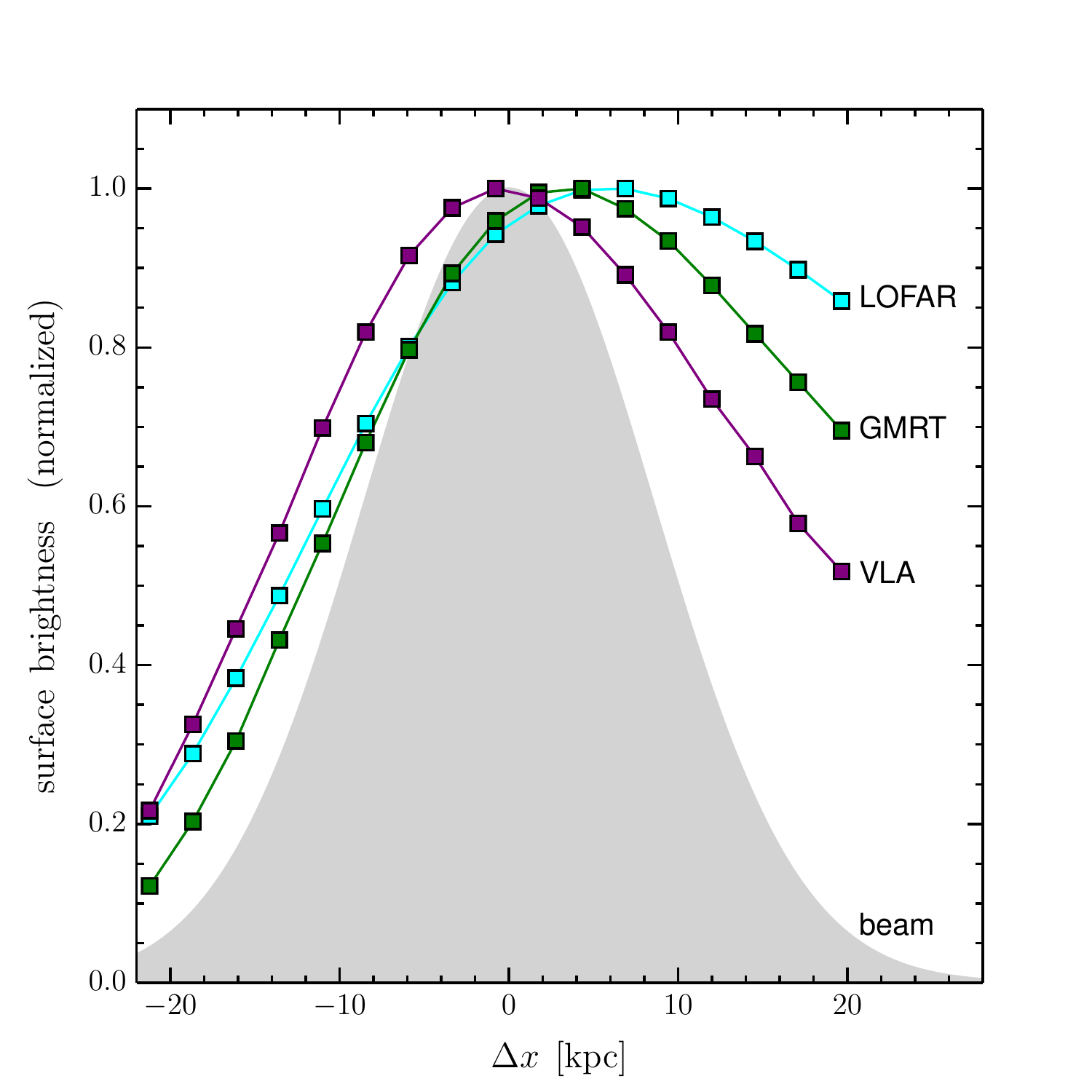}
		\caption{
			Surface brightness profiles measured in the magenta regions shown in Figure~\ref{figure:relicB_spectra} right panel for the VLA, the GMRT and the LOFAR observations, depicted with purple, green, and cyan, respectively. To measure these profiles all observations have been imaged with the same restoring beam width, namely $ 5\farcs5 $, indicated by the grey area. For all three observations, the extracted profiles are normalized by the peak flux. From VLA to LOFAR the peak shifts to the right and gets wider.}
	\end{center}
	\label{fig:peakshift} 
\end{figure}

\begin{figure}
	\begin{center}
		\includegraphics[width=0.45\textwidth]{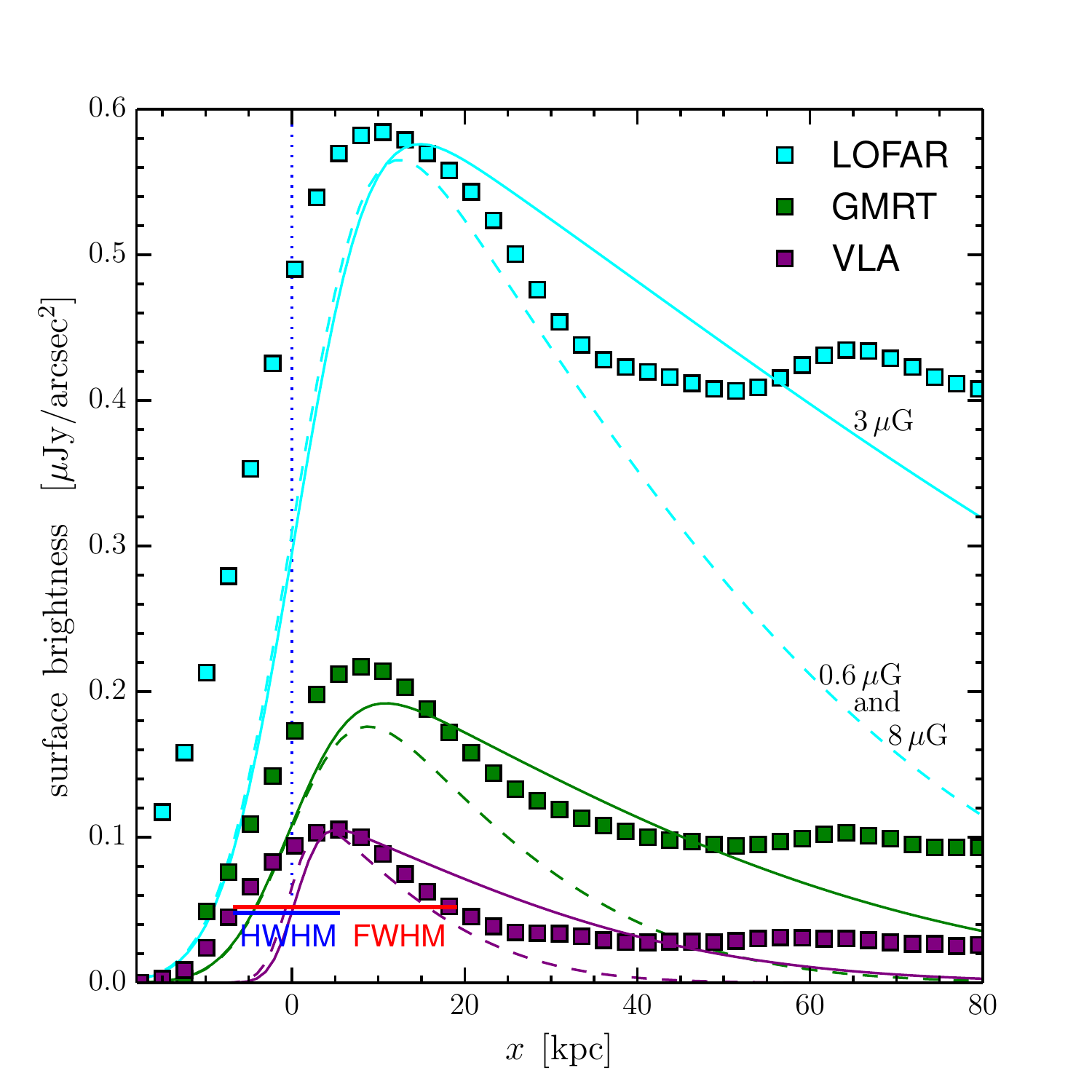}
		\caption{ 
			Surface brightness profiles (squares) measured in the magenta regions shown in Figure~\ref{figure:relicB_spectra} right panel and model profiles assuming a single location of CRe injection at $x=0$ plus subsequent cooling. The resolution of the VLA, GMRT, and LOFAR images are $ 1 \farcs 5 $, $ 3 \farcs 8 $, and  $ 4 \farcs 4 $, respectively. Model profiles are smoothed according to the resolution in the observations and normalized to match the peak surface brightness in the VLA profile.  A Mach number of ${\cal M} = 3.1$ is assumed to match the LOFAR peak surface brightness. Three magnetic field strengths have been assumed for the downstream CRe cooling, i.e. $3\,\rm \mu$ (solid line), $6 \,\rm \mu G$ and $8\,\rm \mu G$ (dashed line). A magnetic field of $3\,\rm \mu G$ would cause a too wide downstream region. The VLA profile shows a rising slope with a half width half maximum of ${\rm HWHM} = 11.8\,\rm kpc$ and width of ${\rm FWHM} = 25.3\,\rm kpc$. }
	\end{center}
	\label{fig:WxDelta} 
\end{figure}

\begin{figure*}
	\begin{center}
		\vspace{0.5cm}
		\hspace{-1.8cm}
		\includegraphics[width=.80\textwidth]{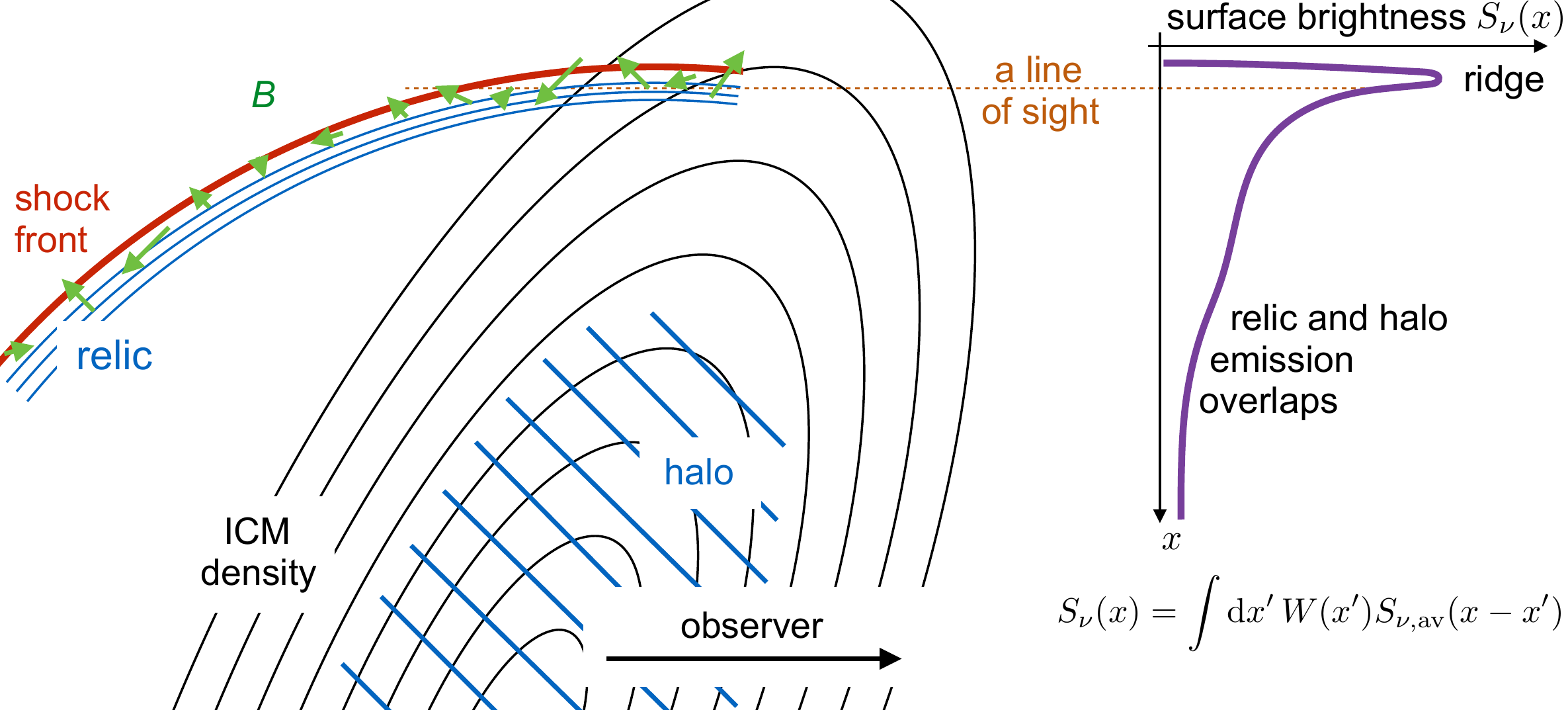}
		\hspace{-1.8cm}
		
		\caption{ 
			Sketch of the proposed shock geometry. The shock front extends behind the cluster. However, the strong depolarization suggests that the emission lies behind the ICM \citep{Kierdorf2016}. With red we indicate the shock front where electrons are accelerated and injected as CRe into the ICM. The green arrows represent a tangled magnetic field in the ICM with varying field strength. The single location of CRe injection at $x=0$ is shown with dotted line. The function $W(x)$ encodes the pile-up of injection at position $x$ due to projection effects. } 
	\end{center}
	\label{fig:ProjektionSketch} 
\end{figure*}

We extract the surface brightness profiles from VLA, GMRT, and LOFAR data in a region where the ridge is particularly narrow, see Figure~\ref{figure:relicB_spectra} right panel. The resultant profiles across the ridge are shown in Fig.~\ref{fig:peakshift}. Evidently, the width and the position of the ridge depend on the observing frequency. The peak surface brightness shifts by about $7\,\rm kpc$ between the VLA and the LOFAR frequencies. The shift is consistent with an intrinsic profile which gets wider at lower frequencies. Such a behavior is expected if the ridge is caused by a shock front seen edge-on and its downstream width is determined by CRe cooling. For a shock front seen perfectly edge-on, the downstream width scales according to $ 1/ \sqrt{\nu_{\rm obs}}$\,, where $\nu\rm_{obs}$ is the observing frequency \citep{Hoeft2007}.


The smallest downstream width has been reported so far for the Sausage-relic, namely $55\:\rm kpc$ at $610 \: \rm MHz$ \citep{vanWeeren2010}. Taking into account that the width scales with the observing frequency, as given above, the ridge is still significantly narrower than the Sausage-relic. The narrowness of the ridge restricts the magnetic field strengths in the downstream area. To quantify this, we compute the downstream profile $S_\nu (x)$ according to \citet{Hoeft2007}. We use a slightly more elaborate computation. For instance, we now include a pitch angle average representing a fast isotropization according to the so-called Jaffe-Perola model \citep{Jaffe1973}. We adopt a power-law electron injection spectrum. Assuming DSA in the test particle regime and a hydrodynamical shock, we can relate the Mach number, the speed of the downstream plasma relative to the shock front, and the spectral index of the CRe distribution \citep{Blandford1987}. Figure~\ref{fig:WxDelta} shows the downstream profiles for three different magnetic field strengths. The Mach number has been chosen to approximately match the peak flux ratio between the VLA and LOFAR profiles. It is evident that a $3\,\mu G$ profile decreases too slowly, i.e. it is not consistent with the narrowness of the ridge. With a significantly higher or lower magnetic field strength the downstream profile would roughly match the narrow ridge. 


Another remarkable feature of the ridge is its northern edge. In the highest resolution VLA profile, it has a half width half maximum (HWHM) of about $11.8\,\rm kpc$, see Figure~\ref{fig:WxDelta}. This is a quite sharp edge, still the HWHM is evidently larger than the beam size, namely $1\farcs 5$ corresponding to $5.5\:\rm kpc$. Therefore, the intrinsic surface brightness profile must show a slope to the north instead of a sharp edge. If the slope is caused by diffusion, we can roughly estimate the diffusion constant. Assuming that the shock front propagates with a speed of the order of $1000\:\rm km\,s^{-1}$, it would sweep up a $5\:\rm kpc$ upstream region within $5\:\rm Myr$. Hence, if diffusion causes the slope, it must distribute the CRe in  $\sim\,5\:\rm Myr$ to distance of $\sim\,5\:\rm kpc$ upstream. This corresponds to a diffusion constant of $D_{\rm diff} \approx 1.5 \times 10^{30} \: \rm cm^2 \, s^{-1} $. The derived value is about one order of magnitude larger than the diffusion constant of CRe transport in the halo of NGC\,7462, found by \citet{2016MNRAS.458..332H}, by assuming that CRe visible in synchrotron emission have energies of a few GeV. However, the value roughly corresponds to the maximum turbulent diffusion found by \cite{2012A&A...544A.103V}. Hence, forming the slope by diffusion would require a significantly more efficient diffusive CRe transport than expected. However, the above calculation assumes that the shock surface is completely flat and perfectly aligned along the line of sight.

We suggests that the ridge is formed by projection of a shock front along the line of sight. The possible scenario is sketched in Figure~\ref{fig:ProjektionSketch}. The shock front is curved, still a large fraction of the front is rather parallel to the line of sight and causes the large surface brightness enhancement of the ridge. Very likely the magnetic field strength in the ICM is not homogeneous. Hence, for such a large shock front seen in projection, it is plausible to assume that the magnetic field strength, which determines the downstream width, varies significantly along the line of sight. The  observed profile is an average 
\begin{equation}
S_{\nu, {\rm av}} (x)
\propto
\int {\rm d}  B \: h( B) \:
S_\nu (x;B) 
\end{equation}
where $S_\nu (x;B)$ is the profile for one particular magnetic field strength and $h(B)$ encodes the fractional abundance of the field strength $B$ along the line of sight. We assume that the shock front is much more extended along the line of sight than the downstream width. Hence, we expect that the magnetic field strength may vary significantly. To model this, we adopt a log-normal distribution as a toy model for a complex magnetic field distribution in the downstream region 
\begin{equation}
h( B; B_0, \sigma ) \,
{\rm d} B 
=
\frac{1}{ \sqrt{2 \pi } \sigma B }
\exp 
\left\{
- \frac{ \ln( B / B_0 ) }{2 \sigma^2 }
\right\} \,
{\rm d} B 
\end{equation}
with a width $\sigma$ and central field strength $B_0$.


\begin{figure*}
	\begin{center}
		\includegraphics[width=0.89\textwidth]{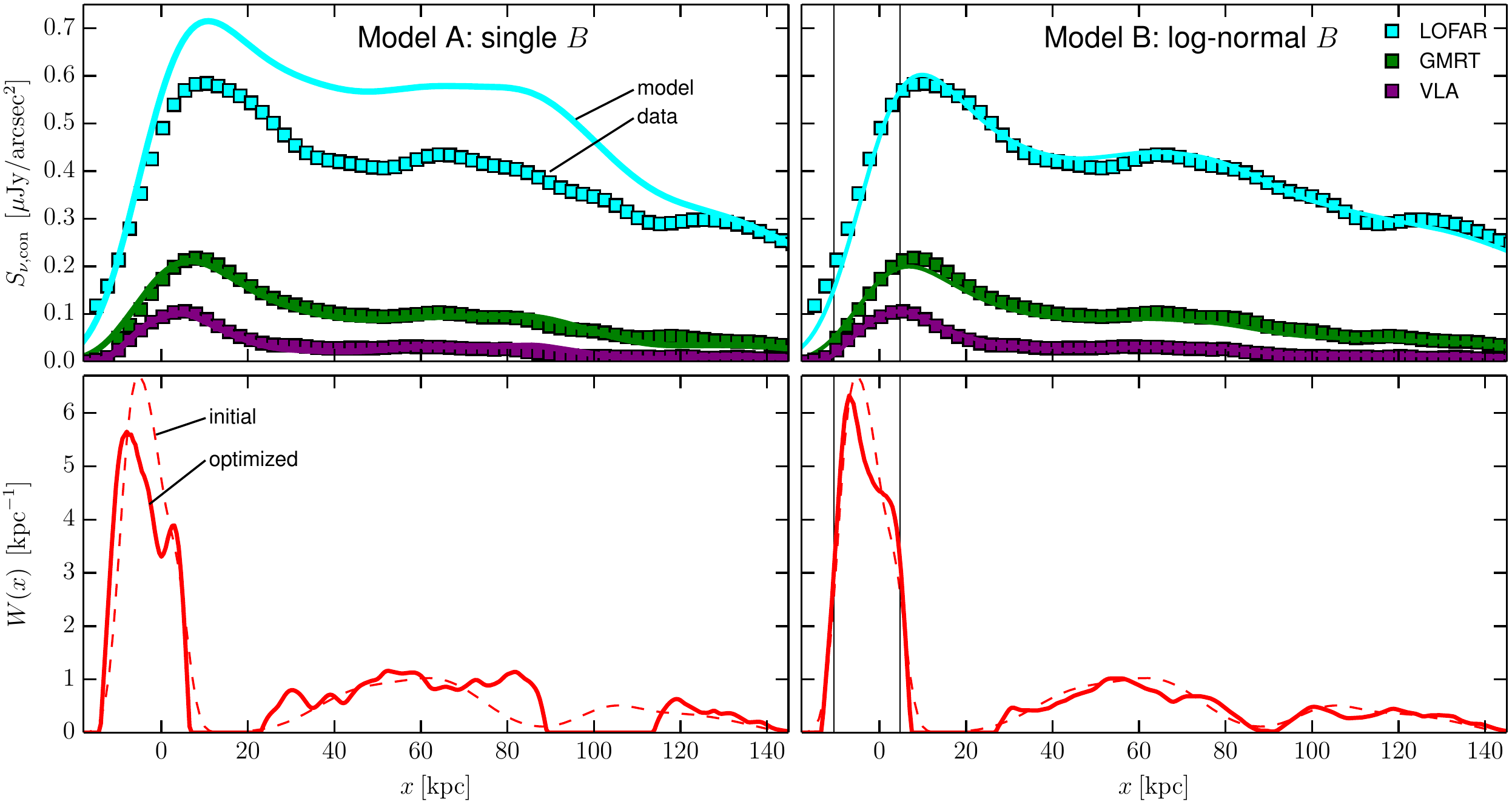}
		\caption{ 
			Surface brightness (top panels) and injection distributions $W(x)$ (bottom panels) for two different models. {\it Left:} A homogeneous magnetic field model; ${\cal M} = 3.0 $, $B_0 = 6.81 \, \mu \rm G$, and $\sigma = 0.001$ (Model A). The lower panel shows the result of the injection distribution optimization, the dashed line shows the initial guess and the solid line the final optimization. {\it Right:} A broad distribution of magnetic field strength; ${\cal M} = 3.75 $, $B_0 = 0.46 \, \mu \rm G$, and $\sigma = 1.3$ (Model B). Model B evidently matches the observed profiles better. The two vertical lines in the right panel indicates that the northern rising slope of the ridge corresponds to an extended region where $W(x)$ is high.}
	\end{center}
	\label{fig:WxOptimize} 
\end{figure*}

The projection of a curved shock front, see Figure~\ref{fig:ProjektionSketch}, also implies that injection not only takes place at the outermost edge but also at what appears to be downstream of the outermost edge. Therefore, the observed surface brightness distribution is the sum of profiles shifted according to the projection and averaged according to the magnetic field distribution. We model this by introducing an injection distribution $W(x)$. Our toy model becomes 
\begin{equation}
S_{\nu, {\rm con}} (x;B_0,\sigma,W)
=
\int
{\rm d} x' \:\:
W(x') 
S_{\nu, {\rm av}} 
(x-x') .
\end{equation} 
We demand here that the sum of the surface brightnesses in all regions $i$ matches the VLA observation 
\begin{eqnarray}
\sum\nolimits_i S_{\nu_0,i}
\: \stackrel{!}{=} \: 
\sum\nolimits_i
S_{\nu_0, {\rm con}}^{\rm mod} (x_i;B_0,\sigma,W)
\end{eqnarray}
where $\nu_0$ indicates the VLA observing frequency. This procedure determines the normalization of the GMRT and LOFAR profiles. This means we do model the spectral index and curvature but we do not model the absolute luminosity. 


\begin{figure*}
	\begin{center}
		\includegraphics[width=0.90\textwidth]{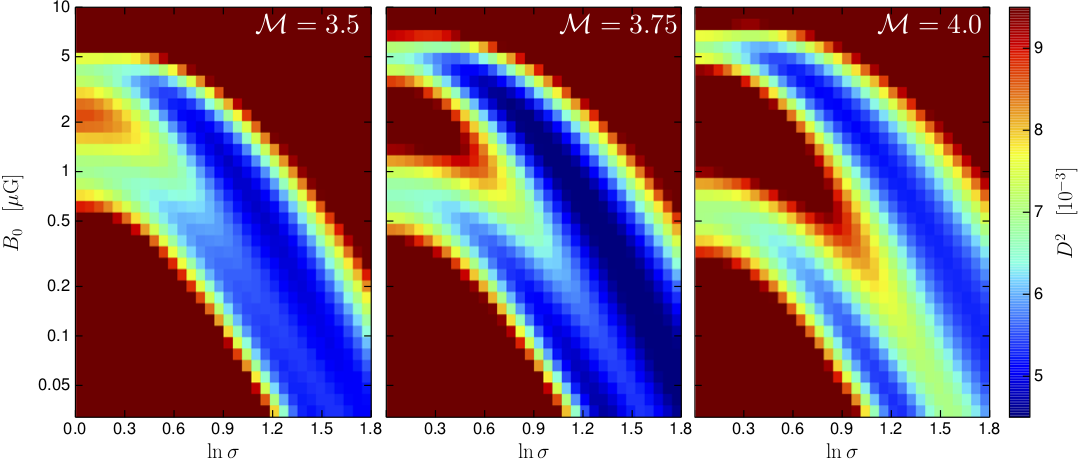}
		\caption{ 
			 Difference $D^2$ between model and observed profiles for different log-normal models for three different Mach numbers. For each $B_0$ - $\sigma$ parameter combination in each panel the injection distribution $W(x)$ has been optimized and the resulting minimum $D^2$ determined. From left to right the model Mach numbers are ${\cal M} = 3.5$, $3.75$, and $4.0$. Dark blue areas indicate where the difference between model and observed profiles is minimal. The Mach number  ${\cal M} = 3.5$ provides the best match. }
	\end{center}
	\label{fig:BsRaster} 
\end{figure*}

We do not know the actual shape of the shock front. Therefore, we estimate the injection distribution from the observed profiles. 
To this end, we measure the difference between model and observed profiles via
\begin{equation}
D^2
=
\frac{1}{S^2_{{\rm max},k}}
\sum\nolimits_k
\frac{1}{N_k}
\sum\nolimits_i
\left(
S_{\nu_k,i}^{\rm obs} 
- 
S_{\nu_k, {\rm con}}^{\rm mod} (x_i;B_0,\sigma,W) 
\right)^2 
\end{equation}
where $S_{\nu_k,i}^{\rm obs}$ denotes the measured surface brightness at frequency $\nu_k$ and position $x_i$, and $S_{\nu_k,i}^{\rm mod}$ the surface brightness at $x_i$ according to our model. We search for the `optimal' injection distribution $W(x)$ using a small routine which varies the injection distribution $W(x)$ by adding and subtracting Gaussians with a random position, width, and amplitude. Modifications which lower $D^2$ are accepted others are rejected. To improve the stability of the optimization scheme, we compute several injection distributions with lower $D^2$ and take the average before proceeding with the optimization. Figure~\ref{fig:WxOptimize} shows the initial guess, the optimized distribution and the resulting surface brightness distribution for two different magnetic field models. For a homogeneous magnetic field, model A, a clear mismatch for the LOFAR profile is evident, despite the fact that the VLA and GMRT match reasonably well. For a broad distribution of magnetic field strengths, model B, we obtain a reasonable match for all observing frequencies.

%
The $W(x)$ which best matches the observation shows several interesting features. First of all, the region where $W(x)$ is high corresponds to the northern slope of the ridge. Therefore, the HWHM found in the slope is determined by the curvature of the shock front or its `mis-alignment' to the line of sight. As a consequence, CRe diffusion must be much less efficient than estimated above. Moreover, we find that there are strong variations of $W(x)$ along the $x$-axis. For instance, we clearly see a double peak in the ridge region, which is in accord with the apparent branching of the ridge to the west, see Figure\,\ref{fig:WxOptimize}. The branching may reflect a complex structure of the actual shock surface. There are also areas where $W(x)$ is close to zero, indicating that the injection varies significantly across the shock front. This may suggest that the small filaments, `bristles', also reflect variations of the injection across the shock front or just variations of the angle between the line of sight and the magnetic field orientation. If the magnetic field is aligned with the line of sight the synchrotron emissivity vanishes. 

%
To estimate which parameters best reproduces the observed profile, we raster the $B_0$ - $\sigma$ parameter space, see Figure~\ref{fig:BsRaster}. Several aspects become evident: magnetic field distributions with $ B_0  \gtrsim  5 \mu G$ do not provide a good match between observed and modeled profiles, irrespective of a narrow or broad distribution of field strengths. For $ B_0  \lesssim 3 \mu G$, there is always a combination of $B_0$ and $\sigma$ which provides a good match, where $\sigma$ becomes larger for lower $B_0$. This may reflect that always some regions are needed which show a wide downstream profile since for about $3\, \rm \mu G$ the downstream profile is widest. Comparing different Mach numbers, we find that ${\cal M} = 3.75$ provides the best match, while for ${\cal M} = 3.5$ and $4.0$ the minimum difference $D^2$ found in the $B_0$ - $\sigma$ parameter space gets larger. The Mach number preferred in the model is even higher than that obtained from flattest region in the spectral index map but agrees with the one obtained from the overall spectrum. In spectral index maps, projection effects and convolution with the beam smear out the profiles and prohibit to actually measure the injection spectral index. Finally, best matches are achieved for a considerably broad magnetic field distribution, namely for $\sigma \gtrsim 0.5$. No uniform magnetic field matches the observations as well as a broad magnetic field distribution.

To summarize, several interesting insights are obtained from the simple toy model. First, one has to be careful when interpreting an observed spectral index as the injection spectrum. In a case of $W(x)$ as wide or wider than the downstream CRe cooling width, projection effects may steepen the observed spectral index i.e. the spectral index can only provide a lower limit for the actual injection spectral index, as shown for the profile investigated here and as found in Section\,\ref{sec:relics_spectra}. For instance, the observed spectral index at the ridge is $\alpha_{\rm inj}\approx-0.70$, but the best match of the profiles is found for a Mach number corresponding to an injection spectral index $\alpha_{\rm inj} = -0.65$. Moreover, our toy model suggests that broad magnetic field distributions are favored with width $\sigma \gtrsim 0.5$ and a central field strength $ B_0 \lesssim 3 \mu G$.


\subsection{Mach numbers derived from integrated spectra, model and spectral index map }
Using VLA data, together with existing LOFAR and GMRT data, we measured an integrated spectral index of $\alpha=-1.15\pm0.02$, suggesting a Mach number of ${\cal M}=3.78^{+0.3}_{-0.2}$, see Table\,\ref{Tabel:Tabel4}.

Our model resulted in a best match for Mach number of ${\cal M}=3.75$ for the northern shock. The integrated spectrum and modeling give consistent Mach numbers.

Another way to obtain a Mach number using radio observations is to measure the injection spectral index directly from the high resolution spectral index maps. From the spectral index map, we find that the flattest spectral index at the northern shock front is about $-0.70$. This injection corresponds to a Mach number of ${\cal M}=3.3^{+0.4}_{-0.3}$. Due to the projection effects discussed in Section\,\ref{sec:downstreamprofiles}, the measured injection index does not necessarily indicate the actual injection spectrum. Hence, the Mach number obtained from the radio spectral index map is a lower limit. 

The deep Chandra observations of 1RXS\,J0603.3+4214 revealed a weak shock of Mach number ${\cal M}_{{{\rm{ X-ray}}}}\approx1.2$ at the northern edge of the Toothbrush \citep{vanWeeren2016}. Clearly, the derived radio and X-ray Mach numbers differ significantly. The recent polarization study of the Toothbrush by \cite{Kierdorf2016} suggests that the relic lies behind the cluster. We argue that the X-ray observation underestimates the strength of the shock since unshocked ICM along the line of sight lowers the measured surface brightness and temperature jumps. We suggest that the shock in the densest region is rather weak, see Figure\,\ref{fig:ProjektionSketch}. Finally, the ridge branching may indicate that the geometry of the shock front is complex and hence the transition might be significantly smeared out. 

\subsection{ Analysis of the halo}
\label{sec:haloanalysis1}

X-ray observations of the cluster 1RXS\,J0603.3+4214 showed that the ICM is dominated by two components \citep{Ogrean2013,vanWeeren2016}. The southern component is brighter than the northern one. The high resolution Chandra image revealed the presence of a cold front at the southern edge of a triangular `bullet-like' structure \citep{vanWeeren2016}. 


\begin{figure}[!thbp]
    \vspace{-0.2cm}
\begin{center}
    \hspace{-3.5cm}
  \includegraphics[width=9.8cm]{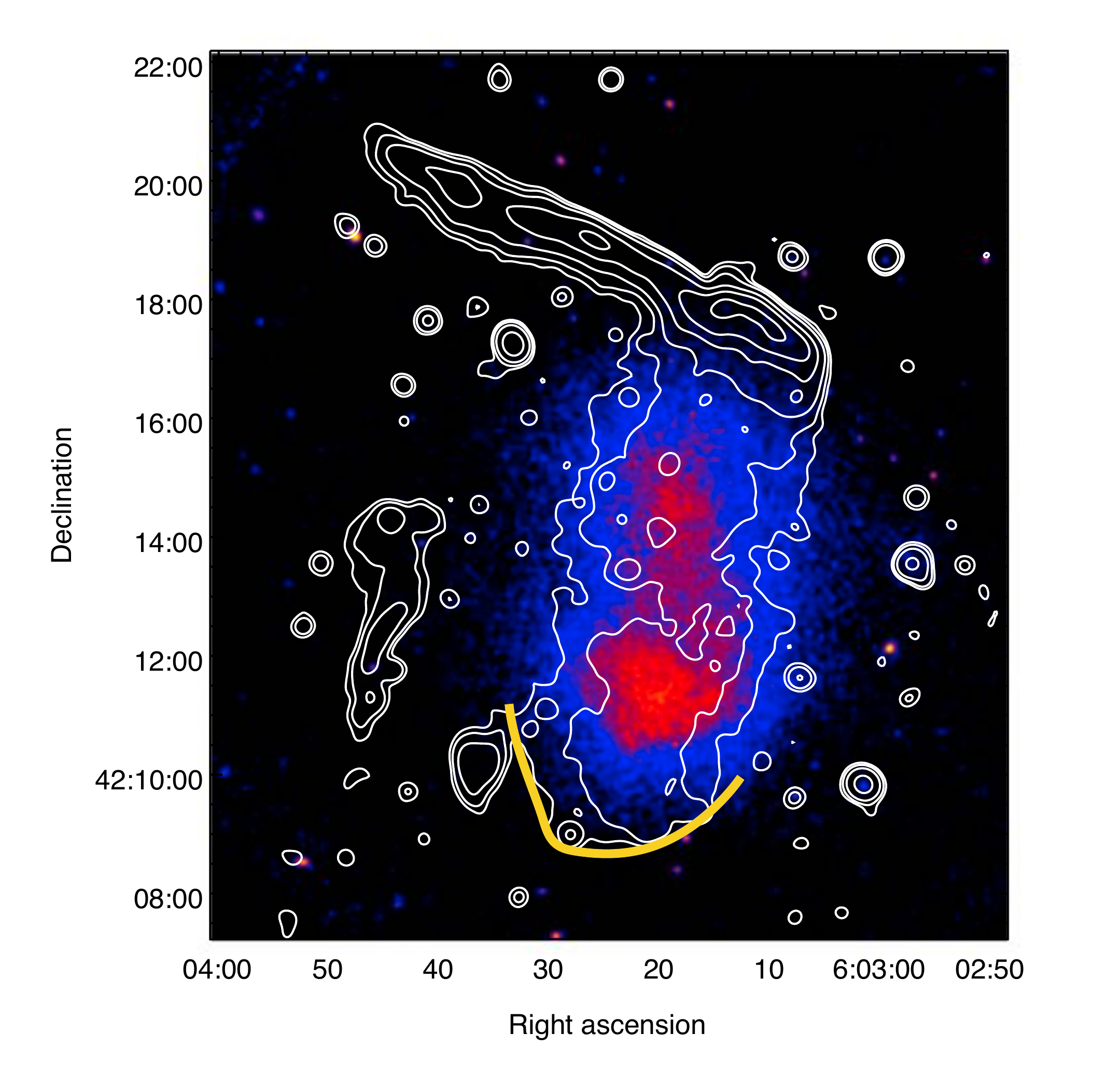}
   \hspace{-3.5cm}
      \vspace{-0.2cm}
\caption{VLA 1-2\,GHz radio contours at the resolution of $15\arcsec$ superposed on the Chandra X-ray image (0.5 to 2 keV band), smoothed to FWHM of $4\arcsec$ \citep{vanWeeren2016}. The VLA image properties are given in Table\,\ref{Tabel:Tabel2}, IM13. Contour levels are drawn at $[1, 2, 4, 8,\dots]\,\times4.5\,\sigma_{{\rm{rms}}}$, where $\sigma_{\rm{rms}}=16\,\mu \rm Jy\,beam^{-1}$. Negative $-4.5\,\sigma_{\rm{rms}}$ contours are shown with dotted lines. The yellow line shows the location of the southern shock front detected in Chandra observations \citep{vanWeeren2016}.}
\end{center}
 \label{figure:xray_radio}
 \end{figure}

In Figure\,\ref{figure:xray_radio}, we compare the X-ray morphology of the cluster to that of the radio emission. The X-ray and the radio morphology are strikingly similar, indicating a connection between the thermal gas and relativistic plasma in this system. The emission from the radio halo extends over the whole region of detected X-ray emission. It is worth emphasizing that the X-ray surface brightness enhancement between the two sub-clusters and the radio surface brightness in the same region follow each other. However, the radio emission extends further to the south where the X-ray emission is fainter, `region S', shown in Figure\,\ref{figure:halocomparison}. It is noteworthy that \cite{vanWeeren2016} recently reported the presence of a shock front (${\hfil \mathcal{M}}=1.39\pm0,06$ via the density jump) at the southern edge of region S, shown with a yellow line in Figure\,\ref{figure:xray_radio}. The southern boundary of region S is edge sharpened and aligns with the southern shock. A similar positioning is observed in others clusters also, for example Bullet cluster \citep{Shimwell2014}, A754 \citep{Macario2011}, A2744 \citep{Owers2011,Pearce2017}, A520 \citep{Vacca2014} and the Coma cluster \citep{Uchida2016}.

\begin{figure}[!thbp]
 \vspace{-0.1cm}
\begin{center}
    \hspace{-3.7cm}
  \includegraphics[width=8.3cm]{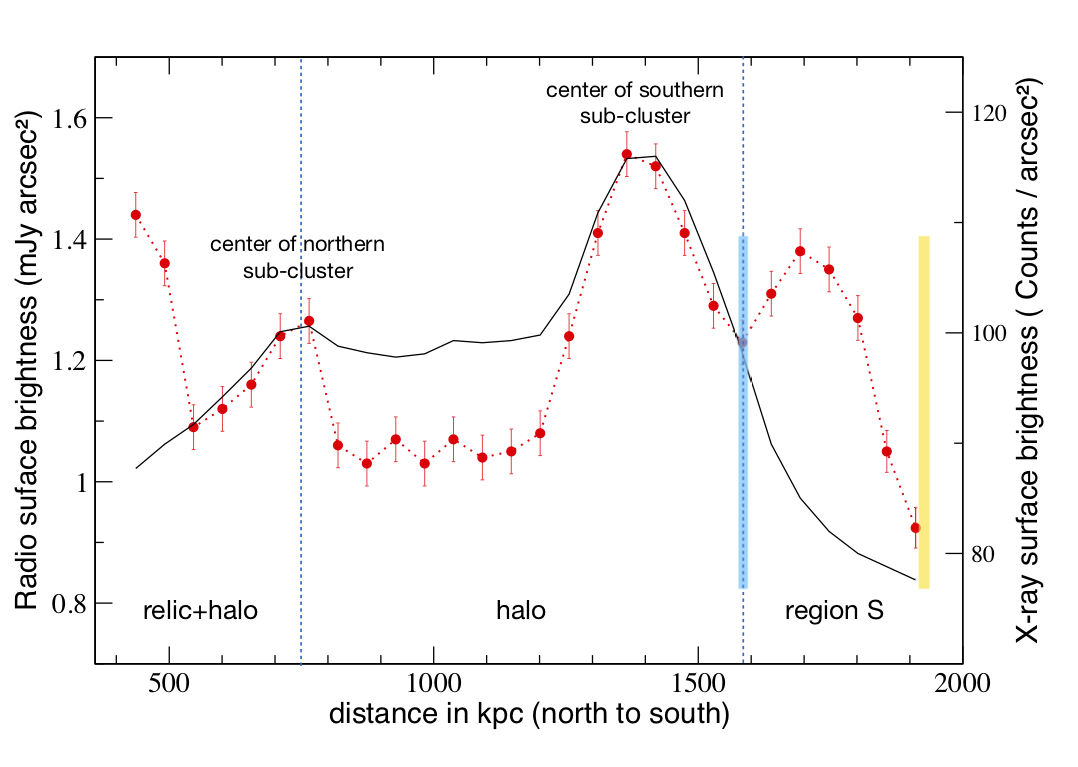}
     \hspace{-4.3cm}
    \vspace{-0.2cm}
  \caption{Radio and X-ray surface brightness profiles along the major axis of the radio halo. The 1.5\,GHz VLA brightness profile at a resolution of $16\arcsec$ is shown with red dots while the Chandra X-ray profile convolved with the same resolution is shown with a black solid line. The VLA image properties are given in Table\,\ref{Tabel:Tabel2}, IM15. The vertical blue and yellow line indicate the location of the cold front and the southern shock front.}
\end{center}
\label{figure:radio_Xray_profiles}
 \end{figure}

In order to examine and compare the X-ray and radio emission going along the line from B1 to region S with distance increasing from north to south, we calculate the average brightness along the major axis of the radio halo. As mentioned in Section\,\ref{sec:halo}, there are 32 discrete sources embedded within the halo. It is important to first subtract discrete unrelated sources. After subtracting strong sources from the uv-data, we measure the surface brightness in the green regions as indicated in Figure\,\ref{figure:integrated} right panel. We smoothed the X-ray image with a Gaussian to a FWHM of 16$\arcsec$, i.e. similar to the restoring beam of our VLA map. The resultant 1.5\,GHz VLA and Chandra profile is shown in Figure\,\ref{figure:radio_Xray_profiles}. The X-ray and radio profile clearly show two sub-cluster components.  We find that the center of the northern and the southern sub-cluster of the radio emission are coincident with the X-ray and there is no offset. However, within the halo there is a region where the radio brightness is relatively lower than the X-ray. We note that the halo region where the radio brightness is lower is dominated by strong discrete sources. We subtracted strong sources, but the source subtraction leaves zero flux at their location which might be a possible reason for the observed low radio brightness. The radio profile also reveals a region of prominent enhanced emission which does not correlate with the X-ray, exactly between the cold front and the southern shock front.

The comparison of the X-ray and the radio brightness profile, as shown in Figure\,\ref{figure:radio_Xray_profiles} suggests that the emission in region S has a different origin than the halo. A cluster undergoing a merger drives shocks and generates turbulence throughout the ICM, providing potential acceleration sites for relativistic particles responsible for radio halos \citep{Feretti2012,Brunetti2014}. As there is a shock detected at the southern edge of region S, therefore, this region could be related to that shock front. We will investigate this possibility in Section\,\ref{sec:halo_spectralindex}.

   \begin{figure}[!thbp]
\vspace{-0.3cm}
\begin{center}
\hspace{-1.0cm}
    \includegraphics[width=9.5cm]{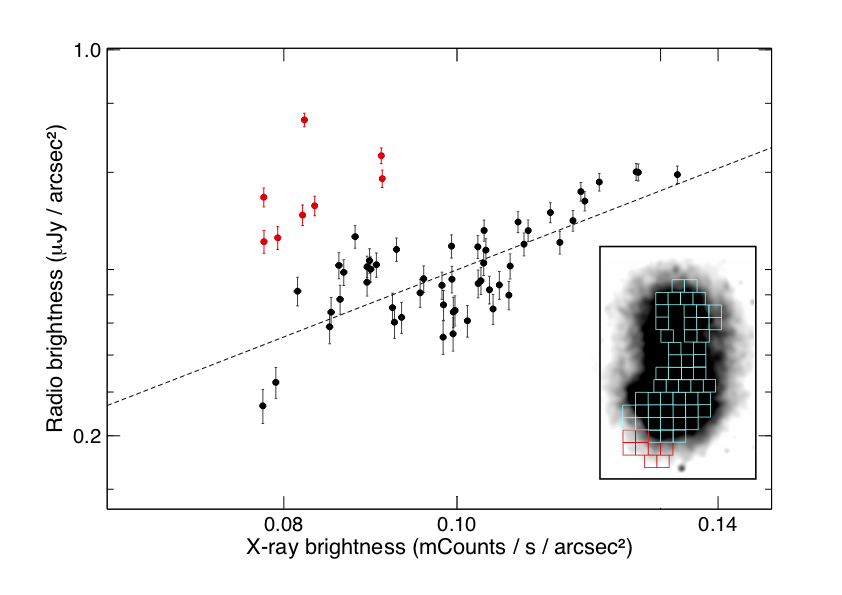}
    \hspace{-1.5cm}
\vspace{-0.4cm}
  \caption{$I_{\rm radio}-I_{\rm\,X-ray}$ relation of the 1RXS\,J0603.3+4214 cluster.  The error-bars represent the RMS of the brightness distribution within each cell. The power-law fit is indicated by a dashed line. Red and cyan boxes, overlaid on the Chandra X-ray image, show the regions where the radio and X-ray surface brightness were extracted. We exclude regions where there is a point source. The surface brightness extracted in cyan cell are indicated by black dots. The red dots are extracted across the southernmost part of the radio halo, i.e. the region between the cold front and the southern shock front, shown with red regions. As discussed in Section\,\ref{sec:haloanalysis1}, the southern part of the radio halo may have a different origin than the rest of the halo.}
\end{center}
\label{figure:halo_correlation}
 \end{figure}

   \begin{figure*}[thbp]
     \vspace{-0.3cm}
\hspace{-1.5cm}
\includegraphics[width=20.5cm]{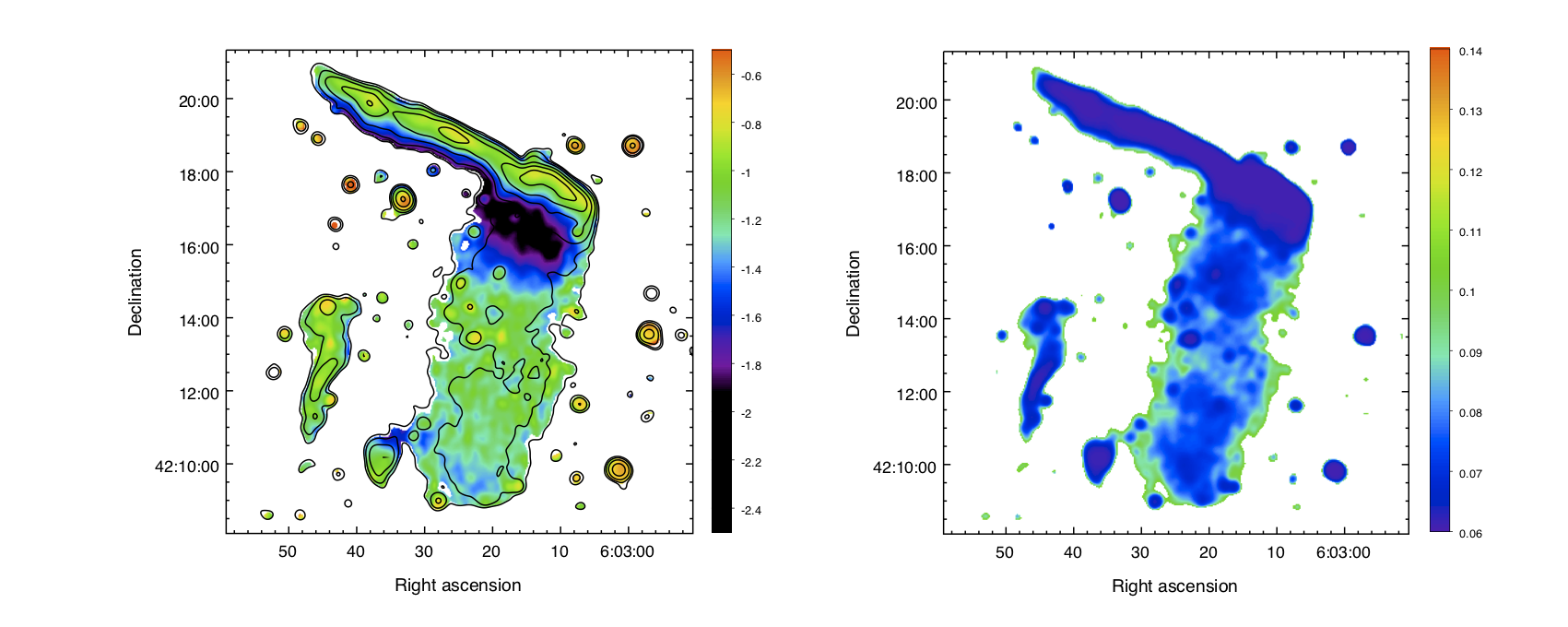}
\hspace{-3.0cm}
\vspace{-0.7cm}
  \caption{{\it Left:} Spectral index map of the radio halo between 150\,MHz and 1.5\,GHz at 16$\arcsec$ resolution. The image properties are given in Table\,\ref{Tabel:Tabel2}, IM14. The contour levels are from the VLA and drawn at $[1, 2, 4, 8, \dots]\,\times4.5\,\sigma_{{\rm{rms}}}$. Negative $-4.5\,\sigma_{\rm{rms}}$ contours are shown with dotted lines. The color bar shows spectral index $\alpha$ from $-2.4$ to $-0.6$. {\it Right:} Corresponding spectral index uncertainty map. The color bar shows the range of values.}
\label{figure:halo_spectral_index_map}
\end{figure*} 

In several clusters hosting a giant radio halo, a radio to X-ray spatial correlation has been observed, e.g. in Abell\,2319, Abell\,2163, Abell\,2744, Abell\,520, Abell\,2255 and the Coma \citep{Govoni2001a,Govoni2001b,Feretti2001,Venturi2013,Vacca2014}. To investigate the possible presence of a radio to X-ray correlation, we performed a quantitative point-to-point comparison of the radio and X-ray brightness. The brightness distribution was constructed using a square grid covering the region where the radio halo emission is higher than the $3\sigma$ level. The grid cell size is equal to $30\arcsec$, corresponding to a physical size of about 109\,kpc. We exclude regions where there is a point source. The plot of the radio versus X-ray brightness, excluding region S, is shown in in Figure\,\ref{figure:halo_correlation}. The radio brightness correlates well with the X-ray brightness: a higher X-ray brightness is associated with a higher radio brightness. The data was fitted with a power law of the type: 
\begin{equation} 
I_{{\rm{{radio}}}} \propto I^{b}_{{\rm{{X-ray}}}}
\end{equation}
where the radio brightness $I_{{\rm{{radio}}}}$ is expressed in $\mu$\,Jy\,beam$^{-1}$ and the X-ray brightness $I_{{\rm{{X-ray}}}}$ in mCounts\,$s^{-1}$arcsec$^{-2}$.
We obtained a slope of $b=$\,1.25$\pm0.16$. This is more or less consistent (within $\,2\,\sigma$) with the radio surface brightness proportional to the X-ray surface brightness, as found for other giant radio halos \citep{Govoni2001a,Govoni2001b}.

The radio and the X-ray brightness distribution across region S, i.e. the region between the cold front and the southern shock front, is indicated by red dots in Figure\,\ref{figure:halo_correlation}. We do not include source W, see Figure\,\ref{figure:Figure1} panel (c) for labeling, as this region consist of two compact sources. The region S is evidently distinct and does not appear to be connected with the rest of the halo.


\subsubsection{Spectral indices}
\label{sec:halo_spectralindex}

The radio halo spectral index maps provide useful information on their origin and connection with the merger processes. The spectral index distribution is also important to test predictions of different models for the electron energy distribution. For example, a systematic variation of the radio halo spectral index with distance from the cluster center is predicted by re-acceleration models \citep{Brunetti2001}. 

The deep VLA images allow us to create spectral index maps for the radio halo at higher resolution and with high accuracy. To make spectral index maps of the halo between 150\,MHz and 1.5\,GHz, we employed the same weighting and uv cut, as mentioned in Section\,\ref{sec:relics_spectra}. We convolve the LOFAR and VLA images to a common resolution of $16\arcsec$.  We did not subtract the extended or point like discrete sources to create the spectral index map but these sources can be well identified from the halo emission. Pixels with flux densities below 5$\sigma_{{\rm{ rms}}}$ were blanked. {Figure\,\ref{figure:halo_spectral_index_map} shows the spectral index image of the radio halo overlaid with the VLA contours. 

The spectral index across the entire halo varies between $-2.0$ and $-0.95$ and shows a slight gradient from north to south. We do not find a region with somewhat steeper spectrum, see Figure\,\ref{figure:halo_spectral_index_map}, at the eastern part of the halo, toward relic E, as found by \cite{vanWeeren2016}. As argued before in Section\,\ref{sec:int_spectrum}, there is a region where, at 1.5\,GHz, the halo dominates the surface brightness and at 150\,MHz the relic dominates. Moreover, the southern part of the halo (region S in Figure\,\ref{figure:halocomparison}) may have a different origin than the central part of the halo. 

The spectral index distribution from north to south, covering the B1 region of the Toothbrush and the entire radio halo is shown in Figure\,\ref{figure:halospectralprofile} . Moving from B1 towards region S, the spectral index varies mainly between $-$0.90 to $-$1.90. The spectrum first steepens in the post-shock regions to $-1.90$ (360\,kpc). After this, the spectrum gradually flattens over a distance of about 400\,kpc (relic+halo region). For a distance of 800\,kpc, the spectrum remains relatively constant with a mean of $-1.16$ (central halo region) and steepens again to the south (region S). 


 \begin{figure}[thbp]
\begin{center}
\hspace{-6.5cm}
\includegraphics[width=9.0cm]{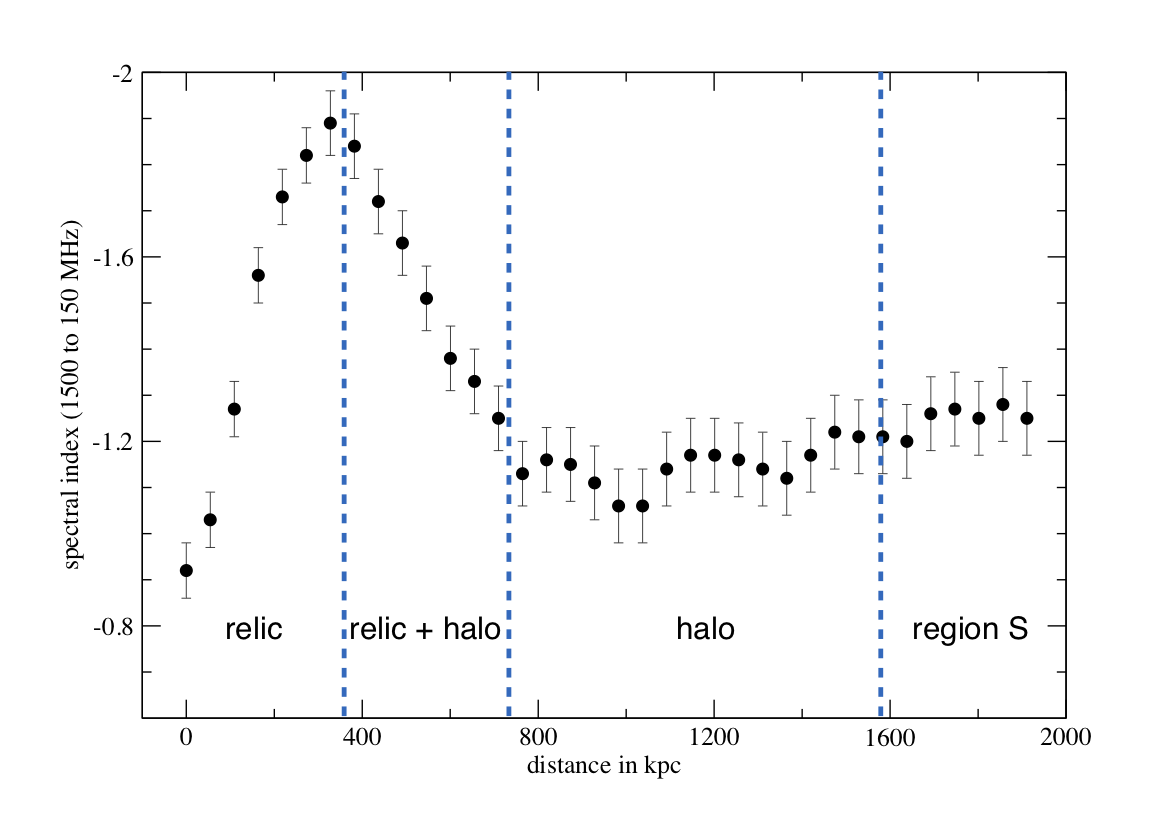}
 \hspace{-6.9cm}
  \vspace{-0.1cm}
 \caption{Extracted spectral index across relic B1 and the radio halo, from north to south. The `relic+halo' indicate combined emission from relic B and the halo. We extract spectral indices in the green regions as indicated in Figure\,\ref{figure:integrated} right panel. Systematic uncertainties in the flux-scale were included in the error bars.}
 \end{center}
\label{figure:halospectralprofile}
\end{figure}

The spectral index of the Toothbrush shows a strong steepening in the downstream areas, while in the relic+halo region, the spectrum flattens progressively. It has been speculated that this spectral behavior indicates a possible connection between shock and turbulence \citep{vanWeeren2016}. 

We note that at 150\,MHz, see Figure\,\ref{figure:halocomparison}, the brush region appears wider with several streams coming out of it, extending from the north to the south. The surface brightness across these streams is very high and the whole emission in this region is mainly dominated by the relic. In contrast, at 1.5\,GHz the same region is dominated by the halo emission. Therefore, the strong steepening and gradual flattening of the spectrum in the `relic+halo' region could be due to the projection of the halo and the relic emission. It has been suggested, that the relic lies behind the cluster \citep{Kierdorf2016}. Hence, the relic and halo could be separated physically, but in projection, one finds the gradual steepening and flattening.

The spectral index across the central part of the halo, excluding the `relic+halo' and region S, is in the range $\,-0.95\leq\alpha\leq-1.25$. To estimate the measurement uncertainties, we followed \cite{Cassano2013}. We find a raw scatter of $0.05$ from the mean spectral index, excluding pixels which are point sources. For the radio halo in 1RXS\,J0603.3+4214, the spectral index uncertainties are much smaller than those of the radio halo in Abell\,665, Abell\,2163, Abell\,520, Abell\,3562, 1E\,0657-55.8, and Abell\,2744 \citep{Feretti2004b,Giacintucci2005,Orru2007,Vacca2014,Shimwell2014}. 

For the radio halo in Abell\,2744, a point-to-point correlation has been observed for the first time, indicating that halo regions with flat spectra correspond to regions with hotter temperatures \citep{Orru2007}. Such a trend is expected, since a fraction of the gravitational energy which is dissipated during cluster merger events in heating thermal plasma, is converted into re-acceleration of highly relativistic particles and amplification of the ICM magnetic field. However, a recent study of the same radio halo by \cite{Pearce2017} reveals that there is no statistically significant correlation. For the present radio halo, the point to point comparison of our high resolution spectral index map to that of the temperature map does not show any correlation. Hence, we confirm that there is no strong correlation between the radio spectral index and X-ray temperature, reported by \cite{vanWeeren2016}. However, the southernmost part of the radio halo, i.e. Region S, is characterized by a hot ICM, namely $T=14$-$15$ keV (Figure 11 of \cite{vanWeeren2016}).


\begin{figure}[!thbp]
\vspace{-0.05cm}
\begin{center}
\hspace{-3.0cm}
    \includegraphics[width=8.0cm]{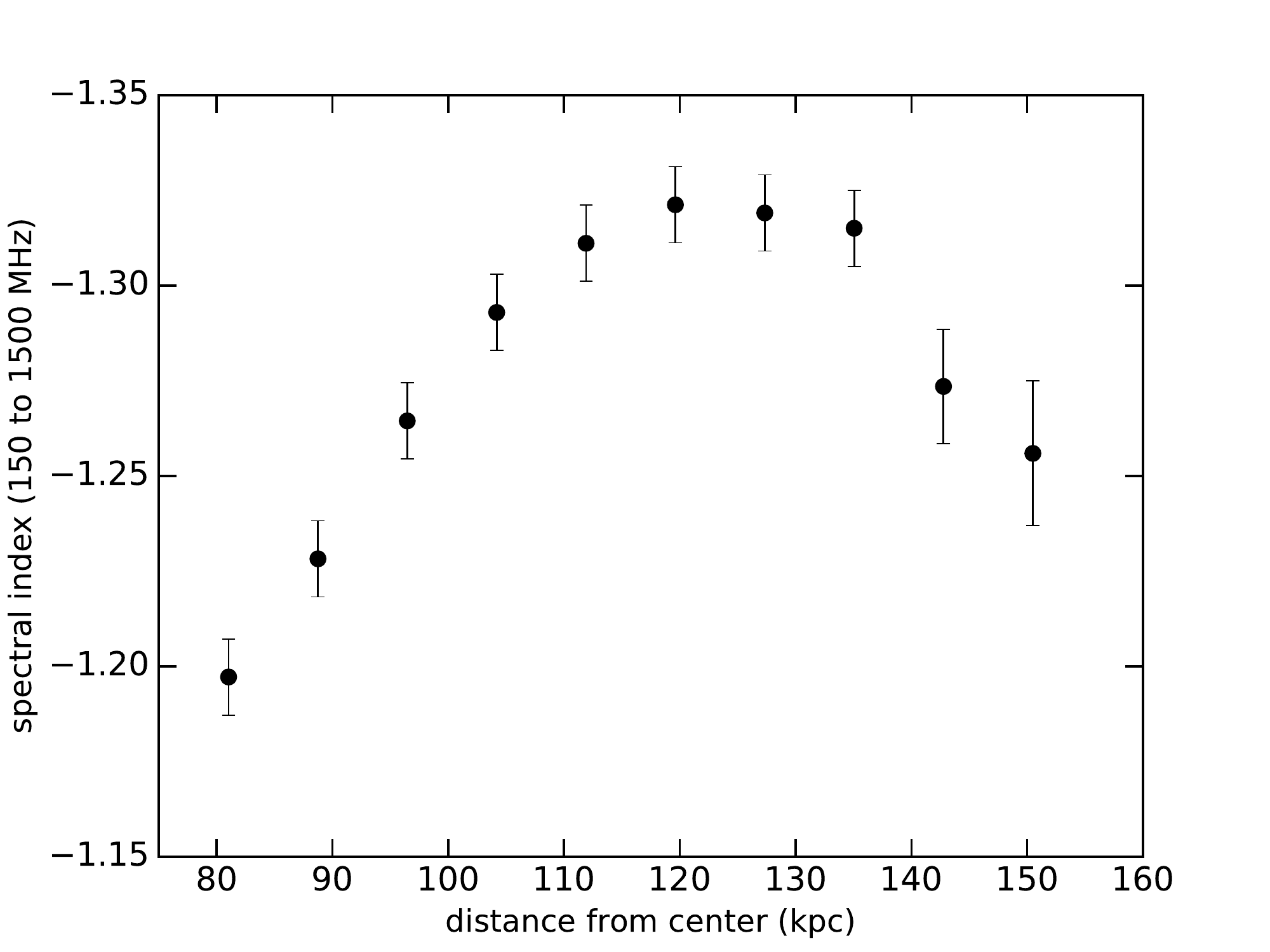}
    \hspace{-3.9cm}
\vspace{-0.1cm}
 \caption{Spectral index distribution across the southernmost part of the radio halo (region S), with distance increasing from north to south. Regions chosen for measuring the radial radio profile is shown in Figure~\ref{figure:integrated} right with red annuli. Systematic uncertainties in the flux-scale were included in the error bars.}
 \end{center}
\label{figure:halo_profile_southern_end}
\end{figure}

To study the spectral distribution across the southernmost region of the halo, i.e. region S, we extract the spectral indices in regions shown with red annuli in Figure\,\ref{figure:integrated} right panel. The resultant distribution is shown in Figure\,\ref{figure:halo_profile_southern_end}. The spectral index distribution in region S shows a clear spectral index gradient, namely $\alpha=-1.20$ to $-1.35$, from north to south.

The southern boundary of the radio halo is relatively well defined and coincides with the southern shock front reported by \cite{vanWeeren2016}. This part of radio halo was speculated to be a fainter relic \citep{vanWeeren2012a} but a uniform spectral index distribution disfavored this scenario. Our spectral index map reveal a spectral index gradient across region S, where the spectrum steepens to $-1.4$ in some areas. In addition, the region S doesn't appear to be associated with the rest of the halo where the radio emission shows clear similarity to the X-ray emission. The total extent of region S at 1.5\,GHz is about 700 kpc. This region is also characterized by a hot ICM as expected for downstream emission of the southern shock front. Considering all evidences, we suggest that the southern part of the radio halo is actually a fainter relic.


\section{Summary and Conclusions}
\label{sec:summary}
We presented deep VLA observations at 1-2\,GHz of the galaxy cluster 1RXS\,J0603.3+4214. These observations were conducted in A, B, C, and D configurations.  Thanks to the wide-band receivers the resulting images show a very low noise level. For a restoring beam width of $1\arcsec$, a noise level of $2\,\mu \rm Jy$ is achieved. Our observations reveal the complex structure of the bright radio relic and confirm a giant radio halo present in the cluster. Below we summarize our primary results:

\begin{itemize}
\item The Toothbrush is evidently made up of filamentary structures. The brush has a striking narrow ridge to its north with a sharp outer edge. The ridge has a width of about $25\,\rm kpc$ and branches to the west. We find several arc-shaped small filaments, `bristles', in the brush region with a width of 3\,to\,5\,\rm kpc that are more or less perpendicular to the ridge. There are two distinct linear filaments, we label them as `double strand', which connect the brush to B2. The B2 region also shows two thin parallel `threads', separated by 5\,kpc.

\item The spectral index map of the Toothbrush between 150\,MHz to 1.5\,GHz shows that at the northern edge of the ridge the spectral index is in the range $-0.70\leq\alpha\leq-0.80$ and steepens within the ridge. The Mach number obtained from the spectral index map is ${\cal M}=3.3^{+0.4}_{-0.3}$. The spectral index across the double strand varies, suggesting a new injection or a change in the local magnetic field. 

\item The surface brightness of the ridge downstream of the shock front at 1-2 GHz decreases by a factor of about $0.5$. We suggest that the surface brightness enhancement of the ridge is caused by a projection effect.

 \item We find that the position of the ridge slightly shifts with observing frequency. This indicates that the intrinsic profile of the emission is frequency dependent as expected for electrons cooling in the downstream region of a shock front seen edge-on. 

\item The VLA observations in combination with published GMRT and LOFAR data allowed us to compare the cooling of the electrons downstream to the shock with a model in which we adopt a significant variation of the magnetic field along the line of sight. The downstream spectral profile can be explained by a log-normal distribution of the magnetic field with a central magnetic field $B_{0}\leq5\,\mu$G, log-normal width $\sigma\geq0.5$, and a Mach number ${\cal M}=3.75$. The model derived Mach number is consistent with the one obtained from the overall spectrum, namely ${\cal M}=3.78^{+0.3}_{-0.2}$. The discrepancy between the X-ray and radio-derived Mach numbers may originate from the fact that the X-ray surface brightness is dominated by the densest region in the ICM along the line of sight where the shock is rather weak. 

\item The fainter relic E comprises three bright, compact regions. These bright regions are evidently not associated with any radio galaxy which could be the source of radio emission. We do not see a significant spectral index steepening across relic E which would be typical for radio relics. 

\item The radio morphology of the large central region of the halo shows a close similarity to the X-ray emitting gas, confirming the connection between the hot gas and relativistic plasma. We find that the radio brightness correlates well with the X-ray brightness in the central region of the halo. The derived power law shows a slope of $b=1.25\pm0.16$, close to a linear correlation. The southernmost region of the halo appears distinct and may have a different origin.

\item The average spectral index across the central region of the radio halo is $\alpha=-1.16\pm0.05$, where the uncertainty gives the raw scatter from the mean spectral index. The spectral index of the halo also shows a slight north to south gradient. The southern part of the halo is steeper and shows a spectral index gradient, suggesting a possible connection with the southern shock detected in Chandra observations.
\item The sensitive and high-resolution radio map allowed us to identify 32 compact radio sources within the radio halo that were not detected previously. These sources, if not subtracted in low resolution radio maps, contribute 25\% of the flux measured for the halo.
\end{itemize}


\textit{Acknowledgments}: We would like to thank the anonymous referee for useful comments. K.R. and M.H. acknowledge support by the research group FOR 1254 funded by the Deutsche Forschungsgemeinschaft: ``Magnetization of interstellar and intergalactic media''. Part of this work was performed at Harvard-Smithsonian Center for Astrophysics. R.J.W. is supported by a Clay Fellowship awarded by the Harvard-Smithsonian Center for Astrophysics. Partial support for L.R. is provided by U.S. National Science Foundation grants 1211595 and 1714205 to the University of Minnesota. F.A-S. acknowledges support from Chandra grant GO3-14131X. This work was performed under the auspices of the U.S. Department of Energy by Lawrence Livermore National Laboratory under Contract DE-AC52-07NA27344. We thank A. Simionescu for helping with Chandra X-ray analysis.

The National Radio  Astronomy Observatory is a facility of the National Science Foundation operated under cooperative agreement by Associated Universities. 

This paper is based (in part) on data obtained with the International LOFAR Telescope (ILT). LOFAR \citep{Haarlem2013} is the Low Frequency Array designed and constructed by ASTRON. It has facilities in several countries, that are owned by various parties (each with their own funding sources), and that are collectively operated by the ILT foundation under a joint scientific policy. We thank the staff of the GMRT that made these observations possible. GMRT is run by the National Centre for Radio Astrophysics of the Tata Institute of Fundamental Research. 
The scientific results reported in this article are based in part on observations made by the Chandra X-ray Observatory and published previously in \cite{vanWeeren2016}. Based in part on data collected at Subaru Telescope, which is operated by the National Astronomical Observatory of Japan.

\bibliography{ref}
\bibliographystyle{apj.bst}

\end{document}